\documentclass[pra,aps,superscriptaddress,noshowpacs,twocolumn,nofootinbib,floatfix]{revtex4}
\usepackage{amsmath,amssymb}
\usepackage{bbold}
\usepackage{graphicx}
\usepackage{hyperref}
\usepackage{doi,url}

\renewcommand{\vec}{\mathbf}
\newcommand{\normord}[1]{{:}\!\mathrel{#1}\!{:}}
\newcommand{\tl}{\tau^\leftarrow}
\newcommand{\tr}{\tau^\rightarrow}
\newcommand{\trs}{\tau^{\rightarrow *}}

\DeclareMathOperator{\arccot}{arccot}
\DeclareMathOperator{\arccoth}{arccoth}
\DeclareMathOperator{\real}{Re}
\begin{document}

\title{Momentum distribution and coherence of a weakly interacting Bose gas after a quench}

\author{Giovanni I. Martone}
\affiliation{LPTMS, UMR 8626, CNRS, Univ. Paris-Sud, Universit\'{e} Paris-Saclay, 91405 Orsay, France}

\author{Pierre-\'{E}lie Larr\'{e}}
\affiliation{Laboratoire de Physique Th\'{e}orique et Mod\'{e}lisation,
Universit\'{e} de Cergy-Pontoise, CNRS, 2 Avenue Adolphe-Chauvin,
95302 Cergy-Pontoise CEDEX, France} 

\author{Alessandro Fabbri}
\affiliation{Departamento de F\'{i}sica Te\'{o}rica and IFIC, Universidad de Valencia-CSIC, \\
C. Dr. Moliner 50, 46100 Burjassot, Spain}
\affiliation{Centro Studi e Ricerche E. Fermi, Piazza del Viminale 1, 00184 Roma, Italy}
\affiliation{Dipartimento di Fisica dell'Universit\`{a} di Bologna and INFN Sezione di Bologna, \\
Via Irnerio 46, 40126 Bologna, Italy}
\affiliation{Laboratoire de Physique Th\'{e}orique, CNRS UMR 8627, B\^{a}t. 210, Univ. Paris-Sud, \\
Universit\'{e} Paris-Saclay, 91405 Orsay Cedex, France }

\author{Nicolas Pavloff}
\affiliation{LPTMS, UMR 8626, CNRS, Univ. Paris-Sud, Universit\'{e} Paris-Saclay, 91405 Orsay, France}

\date{\today}

\begin{abstract}
We consider a weakly interacting uniform atomic Bose gas with a
time-dependent nonlinear coupling constant. By developing a suitable
Bogoliubov treatment we investigate the time evolution of several
observables, including the momentum distribution, the degree of
coherence in the system, and their dependence on dimensionality and
temperature. We rigorously prove that the low-momentum Bogoliubov
modes remain frozen during the whole evolution, while the
high-momentum ones adiabatically follow the change in time of the
interaction strength. At intermediate momenta we point out the
occurrence of oscillations, which are analogous to Sakharov
oscillations. We identify two wide classes of time-dependent
behaviors of the coupling for which an exact solution of the problem
can be found, allowing for an analytic computation of all the
relevant observables. A special emphasis is put on the study of the
coherence property of the system in one spatial dimension. We show
that the system exhibits a smooth ``light-cone effect,'' with
typically no prethermalization.
\end{abstract}

\maketitle

\section{Introduction}
\label{sec:introduction}
A most common manner to study confined ultracold vapors is to remove
the trapping potential and to perform absorption imaging of the
cloud. This is the technique which has been used for imaging the velocity
distribution of the cloud in the first realizations of Bose-Einstein
condensation (BEC) of trapped vapors~\cite{And95,Dav95}. In subsequent
developments of this technique, two-photon Bragg transitions~\cite{Koz99}
have been employed for studying the excitation spectrum
of these systems~\cite{Sta99,Ste02}. In these experiments the momentum
imparted to the condensate is measured by a time-of-flight analysis
after switching off the trapping potential. During the expansion,
quasiparticles turn into real particles which are then imaged (see the
review~\cite{Oze05} where this technique is presented, together with
its extensions). The transformation process at hand is called ``phonon
evaporation'' and has been first theoretically studied in the context of
atomic vapors in Ref.~\cite{Toz04}. It has been shown in this
reference that, at least for cylindrical elongated BECs, the
phenomenon can be effectively described by a quench of the nonlinear
interaction constant whose decrease to zero mimics with good accuracy
the decrease of the mean field experienced by the evaporating
quasiparticle. The rapidity of this quench determines the degree of
adiabaticity of the conversion of a quasiparticle into one (or
several) true particle(s). In more recent and refined experimental
studies, Bragg spectroscopy after a quench in interaction~\cite{Lop17}
and time-of-flight measurements after the opening of the trap~\cite{Cha16}
have been used for studying the momentum distribution and the quantum
depletion.

Actually, this type of quench-like physics is related to a large
variety of physical phenomena and can be envisaged under a number of
different points of view. Only considering the domain of BEC physics
and the related one of quantum fluids of light, it can be used for
studying (1) quantum phonon evaporation~\cite{Toz04,Dui02,Pap16,Fab17},
(2) the dynamics of quantum fluctuations in an expanding BEC and analogies
with cosmology~\cite{Fed04,Uhl05,Uhl09,Ima09,Hun13,Eck18}, (3) dynamical
Casimir effect~\cite{Car10}, (4) correlations~\cite{Jas12} and
entanglement~\cite{Bus14,Rob17a,Rob17b,Gho18,Tia18}, (5) relaxation and
(pre)thermalization~\cite{Che12,Tro12,Gri12a,Lan13,Men15,Lar16,Buc16,Sch18}, 
(6) the Bose gas at unitarity \cite{Yin13,Mak14,Syk14,Eig18},
(7) the degree of coherence in the system, the effects of dimensionality,
the contact parameter~\cite{Cha16,Qu16}, and the condensed
fraction~\cite{Lop17}, (8) formation of jets in two-dimensional
systems~\cite{Cla17}, (9) interaction-induced gauge fields by applying
an interaction strength modulation synchronized with lattice
shaking~\cite{Gol15,Mei16,Fla16,Ple17,Cla18}, and also (10) ``Floquet
engineering'' of the two- and three-body scattering in the
system~\cite{Syk17,Lan18a}.

In this work we present a Bogoliubov treatment of a weakly interacting
uniform BEC system (in any dimension) in which the nonlinear interaction
constant is time dependent. The Bogoliubov method (together with its
low-dimensional generalization) is well documented; it has the drawback
of not taking into account the interaction between quasiparticles, as
recently done in Refs.~\cite{Ber11,Lod12,Zin17,VanR18,Pyl18,Rob18}, and
thus does not address the question of eventual relaxation within the system.
However, the simplicity of the approach makes it possible to present
exact analytic solutions of the problem for wide types of experimentally
relevant time-dependent behaviors. In these cases, one can compute
analytically at each instant of time the momentum distribution of the system
and assess, among many other observables, the degree of coherence in the
system, and its dependence on dimensionality and temperature. In particular,
in a quasi-one-dimensional (1D) regime, we show how the degree of coherence
and the prethermalization are affected by the speed of the quench and
by the initial interaction parameter. We note here that considering a uniform
gas is a simplifying assumption which can be overcome~\cite{Qu16,Sch18}
but already yields interesting qualitative results. Furthermore,
it corresponds to realistic platforms since uniform quantum gases are also
being engineered in the laboratory~\cite{Gau13,Lop17,Eig18}.

The paper is organized as follows. The model and the time-dependent
Bogoliubov treatment are presented in Sec.~\ref{sec:td_Bogo_theory}.
In Sec.~\ref{sec:map_tdho} we show that our system can be mapped onto
an infinite collection of time-dependent harmonic oscillators (TDHOs),
whose properties have been intensively studied in literature.
Section~\ref{sec:solvable_models} is devoted to the analysis of a few
models whose evolution equations can be solved analytically. In these
models, one can compute exactly the one-body density matrix and
characterize the coherence properties, which is done in
Sec.~\ref{sec:coh_prop} for quasi-1D and three-dimensional
(3D) systems. We present our conclusions in Sec.~\ref{sec:conclusion}.
In the appendices we report further information about the change of
the initial time at which the time evolution begins
(Appendix~\ref{sec:change_initial_time}), the proof of the constancy
of the quantum depletion at large times (Appendix~\ref{sec:asymp_depl}),
and some useful properties of hypergeometric functions
(Appendix~\ref{sec:hyperg_func}).

\section{Time-dependent Bogoliubov theory}
\label{sec:td_Bogo_theory}
We start our analysis by summarizing the general features of the Bogoliubov
theory for a BEC with time-dependent coupling constant. We present
two equivalent approaches for studying the time evolution of the system,
which we call the particle and quasiparticle representation. They are the
subject of Secs.~\ref{subsec:part_rep} and~\ref{subsec:quasipart_rep},
respectively. In Sec.~\ref{subsec:exp_values} we show how these two treatments
can be used to calculate the expectation values of the most relevant observables.

\subsection{Particle representation}
\label{subsec:part_rep}
We consider a uniform BEC of $N$ particles of mass $m$ enclosed in a volume
$V$. The particles interact with each other via a two-body repulsive
contact potential, whose $s$-wave scattering length $a(t) > 0$ depends
on time. We assume that the diluteness criterion $\rho a^3(t) \ll 1$,
with $\rho = N/V$ the average density, is fulfilled at any time.
Let $\hat{\Psi}(\vec{r},t)$ denote the atomic field operator, which
depends on time because we choose to work in the Heisenberg
picture. By virtue of the above diluteness condition, we can decompose
this field operator as~\cite{PitStr16}
\begin{equation}
\hat{\Psi}(\vec{r},t) = \psi_0(\vec{r},t) + \delta \hat{\Psi}(\vec{r},t) \, .
\label{eq:gp_split}
\end{equation}
Here, $\psi_0(\vec{r},t)$ is a space- and time-dependent mean field
describing the condensate fraction, whereas $\delta
\hat{\Psi}(\vec{r},t)$ represents the small fluctuations on top of
it. The condensate wave function obeys the Gross-Pitaevskii equation
\begin{equation}
i \hbar \dot{\psi}_0(\vec{r},t)
= - \frac{\hbar^2 \nabla^2}{2m} \psi_0(\vec{r},t)
+ g(t) \left| \psi_0(\vec{r},t) \right|^2 \psi_0(\vec{r},t)
\label{eq:gp_eq}
\end{equation}
and is normalized such that $\int_V d\vec{r} \, \left|
\psi_0(\vec{r},t) \right|^2 = N$. The nonlinear coupling coefficient
is related to the $s$-wave scattering length via $g(t) = 4\pi \hbar^2
a(t) / m$. In writing Eq.~\eqref{eq:gp_eq} we have assumed that no
external trapping is present. As a consequence, if the system is in a
state with uniform density $\rho$ and zero momentum at the initial
time $t_0$, the wave function only acquires a global phase during time
evolution,
\begin{equation}
\psi_0(\vec{r},t) = \sqrt{\rho} \exp[- i\Theta(t)],
\label{eq:gp_sol}
\end{equation}
where
\begin{equation}
\Theta(t)= \int_{t_0}^t d t' \, g(t) \rho / \hbar \, .
\label{eq:gp_phase}
\end{equation}

In order to treat the small fluctuations about the purely condensed
state~\eqref{eq:gp_sol} we shall resort to the Bogoliubov
theory~\cite{PitStr16}. We start by taking the Fourier
expansion of the fluctuation part of the field operator:
\begin{equation}
\delta \hat{\Psi}(\vec{r},t) = \frac{e^{- \displaystyle i\Theta(t)}}{V^{1/2}} 
\sum_{\vec{k} \neq 0}
\hat{a}_{\vec{k}}(t) e^{i \vec{k} \cdot \vec{r}}.
\label{eq:part_Bogo_field}
\end{equation} 
Here, $\hat{a}_{\vec{k}}$
($\hat{a}_{\vec{k}}^\dagger$) are the annihilation (creation) operators
of a particle with momentum $\hbar\vec{k}$.\footnote{Strictly speaking,
the particle operators at time $t$ are $\exp[- i\Theta(t)]
\hat{a}_{\vec{k}}(t)$ and $\exp[i \Theta(t)] \hat{a}_{\vec{k}}^\dagger(t)$.
However, since the phase $\Theta(t)$ plays no role in our calculations,
for brevity we will use the name ``particle operator'' for $a_{\vec{k}}$
and $a_{\vec{k}}^\dagger$.} They obey the standard equal-time bosonic
commutation rules $[\hat{a}_{\vec{k}}(t),\hat{a}_{\vec{k}'}^\dagger(t)] =
\delta_{\vec{k},\vec{k}'}$ and
$[\hat{a}_{\vec{k}}(t),\hat{a}_{\vec{k}'}(t)] =
[\hat{a}_{\vec{k}}^\dagger(t),\hat{a}_{\vec{k}'}^\dagger(t)]=0$.
We henceforth drop hats on operators. The Hamiltonian of the BEC
up to quadratic order in $a_{\vec{k}}$ and $a_{\vec{k}}^\dagger$ is 
\begin{equation}
\begin{split}
H(t) = {} & E_0(t)
+ \sum_{\vec{k} \neq 0} \hbar\Omega_k a_{\vec{k}}^\dagger a_{\vec{k}} \\
&{} \hspace{-8mm} + \frac{g(t) \rho}{2}
\sum_{\vec{k} \neq 0}  
\left(2 a_{\vec{k}}^\dagger a_{\vec{k}} 
+ a_{\vec{k}}^\dagger a_{-\vec{k}}^\dagger
+ a_{\vec{k}} a_{-\vec{k}} + \frac{g(t)\rho}{2\hbar\Omega_k} \right) \, ,
\end{split}
\label{eq:part_Bogo_Ham}
\end{equation}
where $\Omega_k = \hbar k^2/2m$. Aside from the mean-field contribution
$E_0(t) = g(t) N^2 / 2 V$, Hamiltonian~\eqref{eq:part_Bogo_Ham} also
features a constant energy shift proportional to $g^2(t)$. The latter
appears when expressing the coupling constant up to second order
in the scattering length~\cite{PitStr16}. Taken alone,
this term diverges; however, its presence is crucial to ensure the
finiteness of the energy of the instantaneous ground state of the
system (see Sec.~\ref{subsec:quad_rep}).

Before we move on, we point out that, when the interaction
strength is time dependent, the above mean-field and Bogoliubov
approaches are valid if $\tau \gg \tau_{2\mathrm{B}}$. Here, $\tau$ is
the typical time scale characterizing the time variation of $a$,
while $\tau_{2\mathrm{B}} = m a_{\mathrm{max}}^2 / \hbar$, where
$a_{\mathrm{max}} = \max_{t \geq t_0} a(t)$, is the two-body collision
time scale. When $\tau$ is smaller or comparable to $\tau_{2\mathrm{B}}$
(or, more generally, to any few-body time scale), the behavior of
the system becomes sensitive to the presence of a molecular bound
state, which mainly affects the physics of the large-momentum
modes~\cite{Cor15,Cor16,Col18}. The study of such effects is beyond
the scope of this work.

The starting point for studying the time evolution of the system is
represented by the Heisenberg equations for $a_{\vec{k}}$ and
$a_{-\vec{k}}^\dagger$. They form a closed set of first-order
differential equations. By introducing the two-component particle
operator $\vec{A}_{\vec{k}} = (a_{\vec{k}} \,\,
a_{-\vec{k}}^\dagger)^T$, such equations can be cast in a matrix form,
\begin{equation}
i \hbar \dot{\vec{A}}_{\vec{k}}(t)
= \mathcal{H}_{A,k}(t) \vec{A}_{\vec{k}}(t) \, ,
\label{eq:part_evol_eq_mat}
\end{equation}
where
\begin{equation}
\mathcal{H}_{A,k}(t) = 
\begin{pmatrix}
\hbar\Omega_k + g(t)\rho & g(t)\rho \\
-g(t)\rho & -(\hbar\Omega_k + g(t)\rho)
\end{pmatrix} \, .
\label{eq:part_Ham}
\end{equation}
Because of rotational symmetry of the system, all $\vec{A}_{\vec{k}}$'s
with the same $k$ obey the same equation.

The solution of Eq.~\eqref{eq:part_evol_eq_mat} with value
$\vec{A}_{\vec{k}}(t_0)$ at the initial time $t_0$ can be formally
written as
\begin{equation}
\vec{A}_{\vec{k}}(t) = \mathcal{U}_{A,k}(t,t_0) \vec{A}_{\vec{k}}(t_0) \, .
\label{eq:part_evol_sol}
\end{equation}
Here, $\mathcal{U}_{A,k}(t,t_0)$ is a $2 \times 2$ matrix that encodes
the time  evolution of $\vec{A}_{\vec{k}}$ from $t_0$ to $t$. We shall
refer to it as the propagator in the particle representation or, for brevity,
the particle propagator. In order to evaluate $\mathcal{U}_{A,k}(t,t_0)$,
we insert Eq.~\eqref{eq:part_evol_sol} into~\eqref{eq:part_evol_eq_mat},
and we observe that the resulting relation holds for any choice of the initial
value $\vec{A}_{\vec{k}}(t_0)$. This yields
\begin{equation}
i \hbar \dot{\mathcal{U}}_{A,k}(t,t_0) 
= \mathcal{H}_{A,k}(t) \mathcal{U}_{A,k}(t,t_0) \, .
\label{eq:part_evol_eq_prop}
\end{equation}
Thus, the particle propagator is determined by solving the first-order
ordinary differential equation~\eqref{eq:part_evol_eq_prop} with initial
value $\mathcal{U}_{A,k}(t_0,t_0) = \mathbb{1}_{2 \times 2}$.

The particle propagator enjoys the symmetry property
$\sigma_x \mathcal{U}_{A,k}^*(t,t_0) \sigma_x = \mathcal{U}_{A,k}(t,t_0)$,
where $\sigma_x$ is the first Pauli matrix. This follows from the identities
$\mathcal{H}_{A,k}(t) = \mathcal{H}_{A,k}^*(t)$
and $\sigma_x \mathcal{H}_{A,k}(t) \sigma_x = - \mathcal{H}_{A,k}(t)$, together
with the uniqueness of the solution of Eq.~\eqref{eq:part_evol_eq_prop}
with the given initial condition. Thus, it can be written in the form
\begin{equation}
\mathcal{U}_{A,k}(t,t_0) =
\begin{pmatrix}
\alpha_{1,k}(t,t_0) & \alpha_{2,k}^*(t,t_0) \\
\alpha_{2,k}(t,t_0) & \alpha_{1,k}^*(t,t_0)
\end{pmatrix} \, ,
\label{eq:part_evol_op}
\end{equation}
with the two independent complex entries satisfying the initial conditions
$\alpha_{1,k}(t_0,t_0) = 1$ and $\alpha_{2,k}(t_0,t_0) = 0$, as well as
the constraint $|\alpha_{1,k}(t,t_0)|^2 - |\alpha_{2,k}(t,t_0)|^2 = 1$.
The latter ensures the preservation of the equal-time bosonic commutation
rules of the particle operators at all times.

\subsection{Quasiparticle representation}
\label{subsec:quasipart_rep}
In the previous section we have shown how to relate the time evolution
of the system to that of the particle operators. An alternative framework
in which to study the same problem is the quasiparticle representation.
Let us introduce the instantaneous Bogoliubov annihilation [$b_{\vec{k}}(t)$]
and creation [$b_{\vec{k}}^\dagger(t)$] operators through the relation
\begin{equation}
a_{\vec{k}}(t) = u_k(t) b_{\vec{k}}(t) + v_k(t) b_{-\vec{k}}^\dagger(t) \, .
\label{eq:Bogo_trans}
\end{equation}
Here, the instantaneous Bogoliubov weights are given by
\begin{equation}
u_k(t) \pm v_k(t) =  \left[\frac{\Omega_k}{\omega_k(t)}\right]^{\pm 1/2} \, ,
\label{eq:Bogo_uv}
\end{equation}
where $\omega_k(t)$ is the instantaneous Bogoliubov frequency,
\begin{equation}
\hbar\omega_k(t) = \sqrt{\hbar\Omega_k [\hbar\Omega_k + 2 g(t) \rho]} \, .
\label{eq:Bogo_freq}
\end{equation}
Notice that the normalization relation $u_k^2(t) - v_k^2(t) = 1$ holds
at all times, consistently with the preservation of the equal-time bosonic commutation
rules $[b_{\vec{k}}(t),b_{\vec{k}'}^\dagger(t)] = \delta_{\vec{k},\vec{k}'}$ and
$[b_{\vec{k}}(t),b_{\vec{k}'}(t)] = [b_{\vec{k}}^\dagger(t),b_{\vec{k}'}^\dagger(t)]
= 0$ obeyed by the quasiparticle operators. For later convenience, we define
the instantaneous sound velocity
\begin{equation}
c(t) = \sqrt{\frac{g(t) \rho}{m}}
\label{eq:Bogo_sound_vel}
\end{equation}
and the instantaneous condensate healing length
\begin{equation}
\xi(t) = \frac{\hbar}{m c(t)} = \frac{\hbar}{\sqrt{m g(t) \rho}} \, .
\label{eq:Bogo_healing_length}
\end{equation}

By defining the two-component operator $\vec{B}_{\vec{k}} =
(b_{\vec{k}} \,\, b_{-\vec{k}}^\dagger)^T$ we can rewrite the Bogoliubov
transformation~\eqref{eq:Bogo_trans} in matrix form,
\begin{equation}
\vec{A}_{\vec{k}}(t) = 
\mathcal{M}_k(t) \vec{B}_{\vec{k}}(t) \, ,
\label{eq:Bogo_trans_mat}
\end{equation}
where
\begin{equation}
\mathcal{M}_k(t) =
\begin{pmatrix}
u_k(t) & v_k(t) \\
v_k(t) & u_k(t)
\end{pmatrix}
\, .
\label{eq:Bogo_mat}
\end{equation}
The equation governing the time evolution of $\vec{B}_{\vec{k}}$ is
found by inserting Eq.~\eqref{eq:Bogo_trans_mat}
into~\eqref{eq:part_evol_eq_mat} and multiplying on the left by
$\mathcal{M}_k^{-1}(t)$. One obtains
\begin{equation}
i \hbar \dot{\vec{B}}_{\vec{k}}(t) = 
\mathcal{H}_{B,k}(t) \vec{B}_{\vec{k}}(t) \, ,
\label{eq:quasipart_evol_eq_mat}
\end{equation}
with
\begin{equation}
\begin{split}
\mathcal{H}_{B,k}(t) &{}=
\mathcal{M}_k^{-1}(t) \mathcal{H}_{A,k}(t) \mathcal{M}_k(t) \\
& \phantom{={}} - i \hbar \mathcal{M}_k^{-1}(t) \dot{\mathcal{M}}_k(t) \\
&{} =
\hbar
\begin{pmatrix}
\omega_k(t) & 
\displaystyle{\frac{i \dot{\omega}_k(t)}{2\omega_k(t)}} 
\\
\displaystyle{\frac{i\dot{\omega}_k(t)}{2\omega_k(t)}} 
& -\omega_k(t)
\end{pmatrix}
\, .
\end{split}
\label{eq:quasipart_Ham}
\end{equation}
For time-independent $g$, $\mathcal{H}_{B,k}$ is diagonal and
this corresponds to a trivial time evolution of the quasiparticles operators,
that is, $b_{\vec{k}}(t) = \exp[- i \omega_k (t-t_0)] b_{\vec{k}}(t_0)$.
The emergence of the off-diagonal entries is caused by the term
proportional to $\dot{\mathcal{M}}_k(t)$, which in turn appears when
plugging Eq.~\eqref{eq:Bogo_trans_mat} into the left-hand side
of Eq.~\eqref{eq:part_evol_eq_mat} and carrying out the time derivative.
The identity $\dot{u}_k v_k - u_k \dot{v}_k = \dot{\omega}_k / (2 \omega_k)$
has also been used. Physically, the off-diagonal part of
$\mathcal{H}_{B,k}(t)$ is associated with the occurrence of
non-adiabatic effects in the system, being negligible precisely when
the time evolution fulfills the adiabaticity criterion [see
Eq.~\eqref{eq:adiab_cond_gen} below and the related discussion].

The formal solution of Eq.~\eqref{eq:quasipart_evol_eq_mat} is
\begin{equation}
\vec{B}_{\vec{k}}(t) = \mathcal{U}_{B,k}(t,t_0) \vec{B}_{\vec{k}}(t_0) \, .
\label{eq:quasipart_evol_sol}
\end{equation}
The procedure for calculating the quasiparticle propagator
$\mathcal{U}_{B,k}(t,t_0)$ is analogous to that for
$\mathcal{U}_{A,k}(t,t_0)$ (see Sec.~\ref{subsec:part_rep}).
Combining Eqs.~\eqref{eq:quasipart_evol_sol}
and~\eqref{eq:quasipart_evol_eq_mat}, and imposing the result to be
valid for arbitrary $\vec{B}_{\vec{k}}(t_0)$, one gets
\begin{equation}
i \hbar\, \dot{\mathcal{U}}_{B,k}(t,t_0) = 
\mathcal{H}_{B,k}(t) \mathcal{U}_{B,k}(t,t_0) \, .
\label{eq:quasipart_evol_eq_prop}
\end{equation}
This equation has to be solved with initial condition $\mathcal{U}_{B,k}(t_0,t_0)
= \mathbb{1}_{2 \times 2}$.

Similar to the case of the particle propagator, from the property $\sigma_x
\mathcal{H}_{B,k}(t) \sigma_x = - \mathcal{H}_{B,k}^*(t)$ and the uniqueness
of the solution of Eq.~\eqref{eq:quasipart_evol_eq_prop} the identity
$\sigma_x \mathcal{U}_{B,k}(t,t_0) \sigma_x = \mathcal{U}_{B,k}^*(t,t_0)$
follows. This means that the quasiparticle propagator has the form
\begin{equation}
\mathcal{U}_{B,k}(t,t_0) =
\begin{pmatrix}
\beta_{1,k}(t,t_0) & \beta_{2,k}^*(t,t_0) \\
\beta_{2,k}(t,t_0) & \beta_{1,k}^*(t,t_0)
\end{pmatrix} \, ,
\label{eq:quasipart_evol_op}
\end{equation}
where $\beta_{1,k}(t_0,t_0) = 1$, $\beta_{2,k}(t_0,t_0) = 0$, and
$|\beta_{1,k}(t,t_0)|^2 - |\beta_{2,k}(t,t_0)|^2 = 1$ (again, this
is associated with the conservation of the equal-time commutation
rules of the quasiparticle operators).

We conclude the present section by deducing the relationship between
the particle and quasiparticle propagators. For this, we express
$\vec{A}_{\vec{k}}(t)$ and $\vec{A}_{\vec{k}}(t_0)$ in
Eq.~\eqref{eq:part_evol_sol} in terms of $\vec{B}_{\vec{k}}(t)$ and
$\vec{B}_{\vec{k}}(t_0)$ using
Eq.~\eqref{eq:Bogo_trans_mat}. Comparing with
Eq.~\eqref{eq:quasipart_evol_sol} finally yields
\begin{equation}
\mathcal{U}_{B,k}(t,t_0)
= \mathcal{M}_k^{-1}(t) \mathcal{U}_{A,k}(t,t_0) \mathcal{M}_k(t_0) \, .
\label{eq:part_quasipart_evol}
\end{equation}

\subsection{Time evolution of expectation values}
\label{subsec:exp_values}
Let us now see how to employ the formalism introduced in the previous
sections in the study of the time evolution of the expectation values
of the observables. This can be done by applying the standard rule of
the Heisenberg representation: one computes the quantum average of a
given observable at time $t$ over the state of the system at the
initial time $t_0$. Hereby we shall denote this kind of average simply
by $\langle \ldots \rangle$.

The first step is to directly connect the particle operator
$\vec{A}_{\vec{k}}(t)$ at arbitrary time with the quasiparticle
operator $\vec{B}_{\vec{k}}(t_0)$ at the initial time. This can be
accomplished combining either Eq.~\eqref{eq:part_evol_sol} with
Eq.~\eqref{eq:Bogo_trans_mat} at time $t_0$, or
Eq.~\eqref{eq:Bogo_trans_mat} at time $t$ with
Eq.~\eqref{eq:quasipart_evol_sol}. The final result reads as
\begin{equation}
\vec{A}_{\vec{k}}(t) = \mathcal{W}_k(t,t_0) \vec{B}_{\vec{k}}(t_0) \, ,
\label{eq:Bogo_time_trans_mat}
\end{equation}
where we have defined the transformation matrix
\begin{equation}
\begin{split}
\mathcal{W}_k(t,t_0)
&{} = 
\mathcal{U}_{A,k}(t,t_0) \mathcal{M}_k(t_0) \\
&{} = \mathcal{M}_k(t) \mathcal{U}_{B,k}(t,t_0)\\
&{} =
\begin{pmatrix}
U_k(t,t_0) & V^*_k(t,t_0) \\
V_k(t,t_0) & U^*_k(t,t_0)
\end{pmatrix} \, .
\end{split}
\label{eq:Bogo_time_mat}
\end{equation}
The entries of $\mathcal{W}_k(t,t_0)$ are what we denote below as the
``time-propagated'' Bogoliubov weights. Their expressions as functions
of the entries of $\mathcal{U}_{A,k}(t,t_0)$ and $\mathcal{U}_{B,k}(t,t_0)$
are
\begin{subequations}
\label{eq:uv_time}
\begin{align}
\begin{split}
U_k(t,t_0) &{} = u_k(t_0) \alpha_{1,k}(t,t_0) + v_k(t_0) \alpha^*_{2,k}(t,t_0) \\
&{} = u_k(t) \beta_{1,k}(t,t_0) + v_k(t) \beta_{2,k}(t,t_0) \, ,
\end{split}
\label{eq:u_time} \\
\begin{split}
V_k(t,t_0) &{} = u_k(t_0) \alpha_{2,k}(t,t_0) + v_k(t_0) \alpha^*_{1,k}(t,t_0) \\
&{} = u_k(t) \beta_{2,k}(t,t_0) + v_k(t) \beta_{1,k}(t,t_0) \, .
\end{split}
\label{eq:v_time}
\end{align}
\end{subequations}
Notice that $|U_k(t,t_0)|^2 - |V_k(t,t_0)|^2 = 1$ by construction.
The main advantage of using the relation~\eqref{eq:Bogo_time_trans_mat}
[instead of~\eqref{eq:Bogo_trans_mat}] is that it makes it possible
to directly relate the expectation values of the observables to the initial
distribution of quasiparticles. Besides, when written as functions of
$U_k(t,t_0)$ and $V_k(t,t_0)$, these relations retain the same form as in
the case of time-independent coupling, where the weights are given by the
standard Bogoliubov expression.

In this work, we will be mainly interested in the entanglement, the density
fluctuations, and the coherence properties in the system at time $t$.
An important ingredient will be the momentum distribution
\begin{equation}
\begin{split}
n_{\vec{k}}(t) &{} = \langle a_{\vec{k}}^\dagger(t) a_{\vec{k}}(t) \rangle \\
&{} = |V_k(t,t_0)|^2
+ \left[|U_k(t,t_0)|^2 + |V_k(t,t_0)|^2\right] N_{\vec{k}}(t_0) \, ,
\end{split}
\label{eq:depl_uv_time}
\end{equation}
where the quantity
\begin{equation}
N_{\vec{k}}(t_0) = \langle b_{\vec{k}}^\dagger(t_0) b_{\vec{k}}(t_0) \rangle
\label{eq:quasipart_number}
\end{equation}
is the quasiparticle number distribution at the initial time $t_0$.
For obtaining Eq.~\eqref{eq:depl_uv_time} [and also Eqs.~\eqref{eq:str_fac_Bogo}
and~\eqref{eq:dd_corr_uv_time} later in this section] we have
considered an initial state for which the anomalous averages
$\langle b_{-\vec{k}}(t_0) b_{\vec{k}}(t_0) \rangle
= \langle b_{\vec{k}}^\dagger(t_0) b_{-\vec{k}}^\dagger(t_0) \rangle = 0$
(as occurs, for instance, in the case of a thermal state discussed below).

The method we use is able to tackle any type of initial $N_{\vec{k}}(t_0)$.
A usual assumption consists in considering that the initial state corresponds
to a thermal equilibrium at temperature $T$, in which case
\begin{equation}
N_{\vec{k}}(t_0) = \frac{1}{\exp[\hbar\omega_k(t_0)/k_{\mathrm{B}}T] - 1} \, ,
\label{eq:quasipart_number_therm}
\end{equation}
with $k_{\mathrm{B}}$ the Boltzmann constant.
For a system initially in its ground state at zero temperature one has
$N_{\vec{k}}(t_0)= 0$,
\begin{equation}
n_{\vec{k}}(t) = |V_k(t,t_0)|^2 \, ,
\label{eq:depl_uv_time_T0}
\end{equation}
and the initial momentum distribution is given by the standard Bogoliubov
expression
\begin{equation}
n_{\vec{k}}(t_0) = 
\frac{\hbar\Omega_k + g(t_0) \rho}{2\hbar\omega_k(t_0)} - \frac{1}{2} \, .
\label{eq:depl_uv_time_T0_t0}
\end{equation}
One has $n_{\vec{k}}(t_0) \sim m c(t_0) / 2 \hbar k$ for $k \to 0$ and
$n_{\vec{k}}(t_0) \sim \mathcal{C}(t_0)/k^4$ for $k \to + \infty$,
where $\mathcal{C}(t_0) = \xi^{-4}(t_0)$ is the contact parameter at
the initial time.

Other quantities are also important for characterizing the properties
of the system.

\begin{enumerate}
\renewcommand{\theenumi}{\roman{enumi}}
\renewcommand{\labelenumi}{(\theenumi)}
\item The question of entanglement can be addressed by studying the 
four-point correlation function in momentum space:
\begin{equation}
\begin{split}
n^{(2)}_{\vec{k}\vec{k}'}(t) 
= {}&{} \langle \normord{ a_{\vec{k}}^\dagger(t) a_{\vec{k}}(t) 
a_{\vec{k}'}^\dagger(t) a_{\vec{k}'}(t) } \rangle \\
{}&{} - \langle a_{\vec{k}}^\dagger(t) a_{\vec{k}}(t) \rangle 
\langle a_{\vec{k}'}^\dagger(t) a_{\vec{k}'}(t) \rangle \\
= {}&{} n_{\vec{k}}^2(t) \delta_{\vec{k},\vec{k}'} \\
{}&{} + |U_k(t,t_0)|^2 |V_k(t,t_0)|^2 [1 + 2 N_{\vec{k}}(t_0)]^2 
\delta_{\vec{k},-\vec{k}'} \, .
\end{split}
\label{eq:dd_corr_uv_time}
\end{equation}
Here, $\normord{\ldots}$ denotes normal ordering of particle operators.
For a system initially in its ground state at $T=0$ one has
\begin{equation}
n^{(2)}_{\vec{k}\vec{k}'}(t_0) = 
n_{\vec{k}}^2(t_0) \delta_{\vec{k},\vec{k}'}
+ \left[ \frac{g(t_0) \rho}{2\hbar\omega_k(t_0)} \right]^2 
\delta_{\vec{k},-\vec{k}'} \, .
\label{eq:dd_corr_uv_time_T0_t0}
\end{equation}
The quantum non-separability can be tested through the violation of
the Cauchy-Schwarz inequality
\begin{equation}
n^{(2)}_{\vec{k}\vec{k}'}(t) \leq
\sqrt{n^{(2)}_{\vec{k}\vec{k}}(t) n^{(2)}_{\vec{k}'\vec{k}'}(t)} \, .
\label{eq:Cauchy_Schwarz_ineq}
\end{equation}  
In our homogeneous system, such a violation can take place only if
$\vec{k}' = - \vec{k}$ and the following condition is satisfied:
\begin{equation}
n^{(2)}_{\vec{k},-\vec{k}}(t) > n_{\vec{k}}^2(t) \, .
\label{eq:Cauchy_Schwarz_viol}
\end{equation}
\item The question of density fluctuations can be addressed through the
study of the structure factor $S(\vec{k},t)$. The latter is the Fourier
transform of the density-density correlation function or, equivalently,
the regularized Fourier transform of the pair correlation function
(see, e.g.,~\cite{PitStr16}). It can be expressed as
\begin{equation}
S(\vec{k},t) = \frac{1}{N} \sum_{\vec{q},\vec{p}}
\langle a^\dagger_{\vec{q}+\vec{k}}(t) a_{\vec{q}}(t)
a^\dagger_{\vec{p}-\vec{k}}(t) a_{\vec{p}}(t) \rangle \, .
\label{eq:str_fac_def}
\end{equation}
Notice that here the sums are extended over all momenta, including zero.
According to the discussion of Sec.~\ref{subsec:part_rep}, in the presence
of Bose-Einstein condensation one has $a_0(t) = a_0^\dagger(t) = \sqrt{N}$.
Within the accuracy of Bogoliubov theory, the structure factor can be
calculated by retaining in Eq.~\eqref{eq:str_fac_def} only the terms
containing two particle operators with nonvanishing momentum (those with
just one such operator cannot contribute because of momentum conservation).
Then, using the transformation~\eqref{eq:Bogo_time_trans_mat}, one ends up
with
\begin{equation}
S(\vec{k},t) = \frac{\omega_k(t_0)}{\Omega_k}
\left|U_k(t,t_0) + V_k(t,t_0)\right|^2 S(\vec{k},t_0) \, ,
\label{eq:str_fac_Bogo}
\end{equation}
where
\begin{equation}
S(\vec{k},t_0) = \frac{\Omega_k}{\omega_k(t_0)}
\left[2 N_{\vec{k}}(t_0) + 1\right] \, .
\label{eq:str_fac_Bogo_t0}
\end{equation}
\item The coherence properties of the system are intrinsically
related to the degree of Bose-Einstein condensation
(cf. Sec.~\ref{sec:coh_prop}). In 3D, the sum of $n_{\vec{k}}(t)$ over
all nonzero momenta gives the condensate depletion $\Delta N(t)$. By
replacing $\sum_{\vec{k} \neq 0}$ with the integral
$V \int \frac{d^3 k}{(2\pi)^3}$, extended over the whole momentum
space, we can write
\begin{equation}
\Delta N(t) = V \int \frac{d^3 k}{(2\pi)^3} \, n_{\vec{k}}(t) \, .
\label{eq:tot_depl}
\end{equation}
For a system in its ground state at $T=0$ the condensate depletion
\eqref{eq:tot_depl} can be
computed analytically and is given by~\cite{PitStr16}
\begin{equation}
\Delta N(t_0) = \frac{V}{3 \pi^2 \xi^3(t_0)} \, .
\label{eq:tot_depl_T0_t0}
\end{equation}
In two dimensions (2D) and 1D the decomposition~\eqref{eq:gp_split}
of the field operator cannot be performed because the fluctuations of the
phase are not small.  However, quantum fluctuations in reduced dimension
can still be studied within Popov's approach~\cite{Pop72,Pop83}
or, in the case of quasicondensates, through an appropriate extension
of Bogoliubov theory~\cite{Mor03}. In this respect, we point out that
the time-dependent Bogoliubov approach illustrated in this work
is valid in any dimension (see discussions in Refs.~\cite{Mor03,Lar13}).
In Sec.~\ref{sec:coh_prop}
we will use all the above tools to characterize the time evolution
of the one-body density matrix $\rho^{(1)}(\vec{r},\vec{r}',t) =
\langle\hat{\Psi}^\dagger(\vec{r},t) \hat{\Psi}(\vec{r}',t)\rangle$,
which gives information on the coherence properties of the system.
\end{enumerate}

We conclude this section by briefly discussing what happens if the
BEC flows with a finite constant velocity $\vec{v}_0$. In this case,
the condensate wave function is given by the
expression~\eqref{eq:gp_sol} multiplied by the additional phase factor
$\exp\{ i [m\vec{v}_0 \cdot \vec{r} - m v_0^2 (t-t_0)/2] / \hbar\}$.
Concerning the fluctuations on top of the BEC state, the Bogoliubov 
Hamiltonian~\eqref{eq:part_Bogo_Ham} has to be modified adding
the center-of-mass kinetic energy $N m v_0^2/2$ and a further term
$\sum_{\vec{k} \neq 0} \hbar\vec{k} \cdot \vec{v}_0
a_{\vec{k}}^\dagger a_{\vec{k}}$.  This implies the addition of the
quantity $\hbar \vec{k} \cdot \vec{v}_0$ to $\mathcal{H}_{A,k}(t)$ in
Eq.~\eqref{eq:part_Ham} and $\mathcal{H}_{B,k}(t)$ in
Eq.~\eqref{eq:quasipart_Ham}. In turn, the propagators are given by
$\mathcal{U}_{A,k}(t,t_0)$ and $\mathcal{U}_{B,k}(t,t_0)$ calculated
for $\vec{v}_0 = 0$, multiplied by the phase factor $\exp[-i \vec{k}
\cdot \vec{v}_0 (t-t_0)]$. It must be emphasized that, even though this
does not alter the final expression of the observables considered
above, the interpretation of $\hbar\vec{k}$ changes: it represents the
relative momentum of the excitation with respect to the condensate,
the total momentum being $N m \vec{v}_0 + \hbar \vec{k}$.

\section{Mapping onto time-dependent harmonic oscillators}
\label{sec:map_tdho}
The time-dependent Bogoliubov formalism presented in
Sec.~\ref{sec:td_Bogo_theory} is in itself sufficient to fully
determine the time evolution of the relevant observables for any
choice of $g(t)$. However, in most cases the solution of the evolution
equations~\eqref{eq:part_evol_eq_prop}
or~\eqref{eq:quasipart_evol_eq_prop}, that is needed to calculate the
time-propagated Bogoliubov weights~\eqref{eq:uv_time}, can only be
obtained numerically. Even when an analytic solution is available, it
is not always obvious how to determine it. In
Sec.~\ref{subsec:quad_rep} we will show that our problem can be mapped
to the time-dependent harmonic oscillator (TDHO).  This will enable us
to study the properties of the solution in some interesting limiting
cases, such as for low- (Sec.~\ref{subsec:freezing_low_mom}) and
high-momentum modes (Sec.~\ref{subsec:adiab_high_mom}), as well as for
large evolution times (Sec.~\ref{subsec:scatt_form}). This approach
will also prove useful to identify a few exactly solvable models, as
done in Sec.~\ref{sec:solvable_models}.

\subsection{Quadrature representation}
\label{subsec:quad_rep}
We define the quadrature operators as
\begin{subequations}
\label{eq:quad_ops}
\begin{align}
q_{\vec{k}} &{} = \frac{1}{k} (a_{\vec{k}} + a_{-\vec{k}}^\dagger)
= \frac{1}{k} \sqrt{\frac{\Omega_k}{\omega_k(t_0)}} \,
(b_{\vec{k}} + b_{-\vec{k}}^\dagger) \, ,
\label{eq:quad_ops_q} \\
p_{-\vec{k}} &{} = \frac{\hbar k}{2 i} (a_{\vec{k}} - a_{-\vec{k}}^\dagger)
= \frac{\hbar k}{2 i} \sqrt{\frac{\omega_k(t_0)}{\Omega_k}} \,
(b_{\vec{k}} - b_{-\vec{k}}^\dagger) \, .
\label{eq:quad_ops_p}
\end{align}
\end{subequations}
They obey the standard equal-time position-momentum commutation rules
$[q_{\vec{k}}(t),p_{\vec{k}'}(t)] = i \hbar \delta_{\vec{k}\vec{k}'}$,
$[q_{\vec{k}}(t),q_{\vec{k}'}(t)] = 0$, and $[p_{\vec{k}}(t),p_{\vec{k}'}(t)]
= 0$. Furthermore, one has $q_{\vec{k}}^\dagger = q_{-\vec{k}}$
and $p_{\vec{k}}^\dagger = p_{-\vec{k}}$. By rewriting the Bogoliubov
Hamiltonian~\eqref{eq:part_Bogo_Ham} in terms of the quadrature
operators one gets
\begin{equation}
H(t) = E_\mathrm{GS}(t) + 
\sum_{\vec{k} \neq 0} \left[\frac{p_{\vec{k}}^\dagger p_{\vec{k}}}{2m} 
+ \frac{m \omega_k^2(t)}{2} \, q_{\vec{k}}^\dagger q_{\vec{k}}
- \frac{\hbar\omega_k(t)}{2}\right] \, .
\label{eq:quad_Bogo_Ham}
\end{equation}
Equation~\eqref{eq:quad_Bogo_Ham} shows that the system is equivalent
to a collection of infinitely many \textit{complex} harmonic oscillators.
These oscillators are uncoupled and each one is characterized by a
time-dependent angular frequency $\omega_k(t)$. The constant shift
$E_\mathrm{GS}(t)$ is the energy of the instantaneous ground state
of the BEC. The latter is defined as the state that is annihilated
by the expression enclosed in square brackets 
in Eq.~\eqref{eq:quad_Bogo_Ham} at any time $t$ and for any
$\vec{k}$. One has
\begin{equation}
\begin{split}
E_\mathrm{GS}(t) &{} = E_0(t) \\
&\phantom{{}={}} + \frac{1}{2} \sum_{\vec{k} \neq 0}
\left[\hbar\omega_k(t) - g(t)\rho - \hbar\Omega_k
+ \frac{g^2(t)\rho^2}{2 \hbar\Omega_k}\right] \\
&{} = E_0(t) \left[ 1 + \frac{128}{15\sqrt{\pi}} \sqrt{\rho a^3(t)} \right]
\, ,
\end{split}
\label{eq:quad_gs_energy}
\end{equation}
where the second term corresponds to the well-known Lee-Huang-Yang
correction to the mean-field energy of the system~\cite{PitStr16,Lee57}.

The Heisenberg equations for the quadrature operators take the canonical form
\begin{equation}
\dot{q}_{\vec{k}} = \frac{p_{-\vec{k}}}{m} \, , \quad
\dot{p}_{-\vec{k}} = - m \omega_k^2(t) q_{\vec{k}} \, .
\label{eq:quad_evol_eq}
\end{equation}
By deriving the first of Eqs.~\eqref{eq:quad_evol_eq} with respect to time
and combining the result with the second one, we find that $q_{\vec{k}}$
satisfies the TDHO equation
\begin{equation}
\ddot{q}_{\vec{k}} + \omega_k^2(t) q_{\vec{k}} = 0 \, .
\label{eq:quad_tdho_eq_q}
\end{equation}

The solutions of Eqs.~\eqref{eq:quad_evol_eq} with given initial values
$q_{\vec{k}}(t_0)$ and $p_{-\vec{k}}(t_0)$ read as
\begin{subequations}
\label{eq:quad_evol_sol}
\begin{align}
q_{\vec{k}}(t) &{}=
\gamma_{1,k}(t,t_0) q_{\vec{k}}(t_0)
+ \frac{\gamma_{2,k}(t,t_0)}{m \Omega_k} \, p_{-\vec{k}}(t_0) \, ,
\label{eq:quad_evol_sol_q} \\
p_{-\vec{k}}(t) &{}=
m \dot{\gamma}_{1,k}(t,t_0) q_{\vec{k}}(t_0)
+ \frac{\dot{\gamma}_{2,k}(t,t_0)}{\Omega_k} \, p_{-\vec{k}}(t_0) \, .
\label{eq:quad_evol_sol_p}
\end{align}
\end{subequations}
Here, $\gamma_{1,k}$ and $\gamma_{2,k}$ are two real functions.
Inserting Eq.~\eqref{eq:quad_evol_sol_q}
into~\eqref{eq:quad_tdho_eq_q} one immediately verifies that 
$\gamma_{1,k}$ and $\gamma_{2,k}$ both
obey the TDHO equation. The initial conditions they must fulfill are
$\gamma_{1,k}(t_0,t_0) = \dot{\gamma}_{2,k}(t_0,t_0)/\Omega_k = 1$ and
$\gamma_{2,k}(t_0,t_0)/\Omega_k = \dot{\gamma}_{1,k}(t_0,t_0) = 0$.

Let us now express the time-propagated Bogoliubov
weights~\eqref{eq:uv_time} in terms of $\gamma_{1,k}$ and
$\gamma_{2,k}$. To this purpose, we first use Eqs.~\eqref{eq:quad_ops}
to express the left-hand side of Eqs.~\eqref{eq:quad_evol_sol} in
terms of $a_{\vec{k}}(t)$ and $a_{-\vec{k}}^\dagger(t)$, and the
right-hand side in terms of $b_{\vec{k}}(t_0)$ and
$b_{-\vec{k}}^\dagger(t_0)$. Then, we combine the results and compare
with Eqs.~\eqref{eq:Bogo_time_trans_mat}
and~\eqref{eq:Bogo_time_mat}. After a bit of algebra we find
\begin{subequations}
\label{eq:quad_uv_time}
\begin{align}
U_k(t,t_0) + V_k(t,t_0)
&{} = \sqrt{\frac{\Omega_k}{\omega_k(t_0)}} \, \gamma_k(t,t_0) \, ,
\label{eq:quad_uvp_time} \\
U_k(t,t_0) - V_k(t,t_0)
&{} = \sqrt{\frac{\omega_k(t_0)}{\Omega_k}} \, \tilde{\gamma}_k(t,t_0) \, ,
\label{eq:quad_uvm_time}
\end{align}
\end{subequations}
where we have defined
\begin{equation}
\gamma_k(t,t_0) = \gamma_{1,k}(t,t_0)
- i \frac{\omega_k(t_0)}{\Omega_k} \, \gamma_{2,k}(t,t_0)
\label{eq:quad_gamma}
\end{equation}
and
\begin{equation}
\tilde{\gamma}_k(t,t_0) = \frac{i \dot{\gamma}_k(t,t_0)}{\omega_k(t_0)} \, .
\label{eq:quad_gamma_tilde}
\end{equation}
Note that $\gamma_k$ and $\tilde{\gamma}_k$ are connected to the
Fourier transform of the density and phase fluctuations, respectively
(see, e.g., Refs.~\cite{PitStr16,Lar13}). Being a linear combination
of $\gamma_{1,k}$ and $\gamma_{2,k}$, $\gamma_k$ also fulfills
the TDHO equation
\begin{equation}
\ddot{\gamma}_k + \omega_k^2(t) \gamma_k = 0 \, ,
\label{eq:quad_tdho_eq_gamma}
\end{equation}
with initial conditions
\begin{equation}
\gamma_k(t_0,t_0) = 1 \, ,
\quad \dot{\gamma}_k(t_0,t_0) = - i \omega_k(t_0) \, .
\label{eq:quad_tdho_eq_ic}
\end{equation}
Additionally, computing the Wronskian of $\gamma_k$ and
$\gamma_k^*$ one finds
\begin{equation}
\real\left[\tilde{\gamma}_k^*(t,t_0) \gamma_k(t,t_0)\right] = 1 \, .
\label{eq:quad_tdho_curr_cons}
\end{equation}
This automatically ensures that the operators given in
Eqs.~\eqref{eq:quad_evol_sol} satisfy the standard equal-time
position-momentum commutation rules at all times. We note here
for completeness that Eq.~\eqref{eq:str_fac_Bogo} can be rewritten
in a concise form as $S(\vec{k},t) = |\gamma_k(t,t_0)|^2 S(\vec{k},t_0)$.

In conclusion, the whole problem of calculating the time-propagated
Bogoliubov weights is reduced to finding a single solution of
Eq.~\eqref{eq:quad_tdho_eq_gamma}. This result will play a crucial role
in the rest of this paper.

\subsection{Freezing of the low-momentum modes}
\label{subsec:freezing_low_mom}
In this section we prove that the low-$k$ modes are not affected by
the time dependence of $g$, provided that $g(t_0) \neq 0$. In the
regime of $k \ll \min\{\xi^{-1}(t_0),\sqrt{m/\hbar\tau}\}$, where
$\tau$ is the typical time scale characterizing the time variation of $g$,
we can treat the second term on the left-hand side of
Eq.~\eqref{eq:quad_tdho_eq_gamma} as a perturbation. Consequently,
the solution can be expanded as
$\gamma_k(t,t_0) = \sum_{j=0}^{+\infty} \gamma_k^{(j)}(t,t_0)$, where
the superscript denotes the order in $k$. Let us insert this expansion
into Eq.~\eqref{eq:quad_tdho_eq_gamma}, collect all the terms of the
same order in $k$, and equate each one of them to zero.  Up to second
order in $k$ we find
\begin{equation}
\ddot{\gamma}_k^{(0)} = \ddot{\gamma}_k^{(1)} = 0 \, ,
\quad \ddot{\gamma}_k^{(2)} + c^2(t) k^2 \gamma_k^{(0)} = 0 \, .
\label{eq:pert_gamma_eq}
\end{equation}
Recalling that $\omega_k(t_0) = c(t_0) k + O(k^3)$, from the initial
conditions~\eqref{eq:quad_tdho_eq_ic} for $\gamma_k$ one immediately
finds those for the $\gamma_k^{(j)}$ up to $j=2$: $\gamma_k^{(0)}(t_0,t_0)
= 1$, $\gamma_k^{(1)}(t_0,t_0) = \gamma_k^{(2)}(t_0,t_0) = 0$,
$\dot{\gamma}_k^{(0)}(t_0,t_0) = \dot{\gamma}_k^{(2)}(t_0,t_0) = 0$,
$\dot{\gamma}_k^{(1)}(t_0,t_0) = - i c(t_0) k$. The integration of
Eqs.~\eqref{eq:pert_gamma_eq} is straightforward, and the final result
for $\gamma_k$ is
\begin{equation}
\begin{split}
\gamma_k(t,t_0) = {}&{} 1 - i c(t_0) (t-t_0) k \\
&{} - \int_{t_0}^t d t' \int_{t_0}^{t'} d t'' c^2(t'') k^2 + O(k^3) \, .
\end{split}
\label{eq:pert_gamma_sol_g0f}
\end{equation}
This yields the low-$k$ behavior of the the time-propagated
weights~\eqref{eq:quad_uv_time} under the form
\begin{subequations}
\label{eq:pert_uv_time_g0f}
\begin{align}
U_k(t,t_0) &{} = \frac{1}{2} \sqrt{\frac{2 m c(t_0)}{\hbar k}}
\left[ 1 + \frac{Z(t,t_0) \hbar k}{2 m c(t_0)} + O(k^2) \right] \, ,
\label{eq:pert_u_time_g0f} \\
V_k(t,t_0) &{} = \frac{1}{2} \sqrt{\frac{2 m c(t_0)}{\hbar k}}
\left[ -1 + \frac{Z^*(t,t_0) \hbar k}{2 m c(t_0)} + O(k^2) \right] \, ,
\label{eq:pert_v_time_g0f}
\end{align}
\end{subequations}
where
\begin{equation}
Z(t,t_0)= 1 - 2 i \int_{t_0}^t d t' \, m c^2(t') / \hbar \, .
\label{eq:pert_z}
\end{equation}
Equations~\eqref{eq:pert_uv_time_g0f} show that the leading-order
term in the low-$k$ expansion of the time-propagated Bogoliubov
weights is independent of time. Therefore, for $k \to 0$ all the
$k$-dependent observables remain frozen to their initial values
during the time evolution, irrespective of the specific form of $g(t)$.

It is instructive to see what happens if $g(t_0) = 0$. In this case
one can still perform the low-$k$ expansion leading to
Eqs.~\eqref{eq:pert_gamma_eq}. The only change concerns
the initial values of the time derivatives of the $\gamma_k^{(j)}$'s:
$\dot{\gamma}_k^{(0)}(t_0,t_0) = \dot{\gamma}_k^{(1)}(t_0,t_0) = 0$,
$\dot{\gamma}_k^{(2)}(t_0,t_0) = - i \Omega_k$. Then,
Eq.~\eqref{eq:pert_gamma_sol_g0f} is replaced by
\begin{equation}
\gamma_k(t,t_0)
= 1 - i \int_{t_0}^t d t' Z(t',t_0) \frac{\hbar k^2}{2m} + O(k^4) \, .
\label{eq:pert_gamma_sol_g00}
\end{equation}
The final result for the time-propagated weights~\eqref{eq:quad_uv_time}
is, up to leading order in $k$,
\begin{subequations}
\label{eq:pert_uv_time_g00}
\begin{align}
U_k(t,t_0) &{} = 1 - i \int_{t_0}^t d t' \, m c^2(t') / \hbar + O(k^2) \, ,
\label{eq:pert_u_time_g00} \\
V_k(t,t_0) &{} = i \int_{t_0}^t d t' \, m c^2(t') / \hbar + O(k^2) \, .
\label{eq:pert_v_time_g00}
\end{align}
\end{subequations}
Thus, the low-$k$ modes also evolve in time if the coupling constant
is initially vanishing.

Finally, one should notice that the time-dependent terms in
Eqs.~\eqref{eq:pert_uv_time_g0f} and~\eqref{eq:pert_uv_time_g00}
may diverge as $t \to + \infty$, i.e., they may be secular terms. This
is just an artifact of the perturbative expansion carried out in this
section. The correct large-$t$ behavior of all quantities can be
obtained through the inclusion of the higher-order terms in $k$.
However, the conclusion that the low-$k$ modes are frozen during
the time evolution if $g(t_0) \neq 0$ only requires the constancy
of the leading term of Eqs.~\eqref{eq:pert_uv_time_g0f}. Thus,
it holds irrespective of the behavior of the subleading contributions.

\subsection{Adiabatic behavior of high-momentum modes}
\label{subsec:adiab_high_mom}
Let us now move to the study of the high-$k$ modes. As we shall prove,
such modes are able to adiabatically follow the time dependence of the
nonlinear coupling coefficient. For this purpose, we write the complex
function $\gamma_k$ in terms of two real quantities $A_k$ and $S_k$,
corresponding to its amplitude and phase degrees of freedom,
respectively~\cite{Kul57},
\begin{equation}
\gamma_k(t,t_0) = A_k(t,t_0) e^{i S_k(t,t_0)} \, .
\label{eq:adiab_gamma_amp_phase}
\end{equation}
From Eq.~\eqref{eq:quad_tdho_eq_ic} we find that the amplitude and phase
must obey the initial conditions $A_k(t_0,t_0) = 1$, $\dot{A}_k(t_0,t_0) = 0$,
$S_k(t_0,t_0) = 0$, and $\dot{S}_k(t_0,t_0) = - \omega_k(t_0)$. Inserting
Eq.~\eqref{eq:adiab_gamma_amp_phase} into Eq.~\eqref{eq:quad_tdho_eq_gamma},
and separating the real and imaginary parts, one gets the coupled second-order
equations
\begin{subequations}
\label{eq:adiab_amp_phase_eq}
\begin{align}
&{} \ddot{A}_k - A_k \dot{S}_k^2 + \omega_k^2(t) A_k = 0 \, ,
\label{eq:adiab_amp_eq} \\
&{} A_k \ddot{S}_k + 2 \dot{A}_k \dot{S}_k = 0 \, .
\label{eq:adiab_phase_eq}
\end{align}
\end{subequations}
Equation~\eqref{eq:adiab_phase_eq} can be 
integrated straightforwardly. This gives, taking the
initial conditions into account,
\begin{equation}
\dot{S}_k = - \omega_k(t_0) A_k^{-2} \, .
\label{eq:adiab_phase_der_sol}
\end{equation}
Substituting Eq.~\eqref{eq:adiab_phase_der_sol}
into~\eqref{eq:adiab_amp_eq} yields a nonlinear equation for the
sole $A_k$~\cite{Lew67}, sometimes called the Ermakov-Pinney-Milne
(EPM) equation:
\begin{equation}
\ddot{A}_k + \omega_k^2(t) A_k = \omega_k^2(t_0) A_k^{-3} \, .
\label{eq:adiab_epm_eq}
\end{equation}
All the calculations up to this point are exact. However, the EPM
equation is usually hard to solve, except for some specific choices of
the time-dependent coupling $g(t)$. Here, we are interested in finding
an approximate solution in the limit where the time evolution of the
system is slow (adiabatic). For a TDHO, this happens when the temporal
variation of the frequency $\omega_k$ occurs on a time scale
$\tau$ much larger than the instantaneous oscillation period
$2\pi/\omega_k(t)$ at any time (a more quantitative adiabaticity
criterion is provided below). If this holds, the second-order derivative
$\ddot{A}_k$, being proportional to $\tau^{-2}$ (recall that all quantities
depend on $t/\tau$), can be neglected with respect to the other term
$\omega_k^2(t) A_k$ in the left-hand side of Eq.~\eqref{eq:adiab_epm_eq}.
One finds
\begin{equation}
A_k(t,t_0) = \sqrt{\frac{\omega_k(t_0)}{\omega_k(t)}} \, .
\label{eq:adiab_amp_sol}
\end{equation}
Then, integration of Eq.~\eqref{eq:adiab_phase_der_sol} immediately
yields $S_k$. The final result for the adiabatic solution
of the TDHO equation~\eqref{eq:quad_tdho_eq_gamma} is
\begin{equation}
\gamma_k(t,t_0)
= \sqrt{\frac{\omega_k(t_0)}{\omega_k(t)}} \,
\exp\left[- i \int_{t_0}^t d t' \, \omega_k(t')\right] \, .
\label{eq:adiab_gamma_sol}
\end{equation}
The corresponding time-propagated Bogoliubov weights are
obtained from Eqs.~\eqref{eq:quad_uv_time}. Neglecting
terms proportional to $\dot{\omega}_k(t)/\omega_k^2(t)$
[see Eq.~\eqref{eq:adiab_cond_gen} below] one has
\begin{subequations}
\label{eq:adiab_uv_time}
\begin{align}
U_k(t,t_0) &= u_k(t) \exp\left[- i \int_{t_0}^t d t' \, 
\omega_k(t')\right] \, ,
\label{eq:adiab_u_time} \\
V_k(t,t_0) &= v_k(t) \exp\left[- i \int_{t_0}^t d t' \, 
\omega_k(t')\right] \, .
\label{eq:adiab_v_time}
\end{align}
\end{subequations}
Thus, in this case the time-propagated weights
coincide with the instantaneous ones, up to a global dynamic phase
factor. This means that all the observables for a BEC with a
time-dependent coupling have the same expression as for a static
condensate, with the static coupling replaced by $g(t)$: the evolution 
is adiabatic.

It remains to better clarify the conditions of validity of the
adiabatic approach. As mentioned above, it requires
$\omega_k^2(t) A_k(t) \gg |\ddot{A}_k(t)|$. Taking
expression~\eqref{eq:adiab_amp_sol} for $A_k(t)$, and assuming
the two characteristic times
$\left|\dot{\omega}_k(t)\right| /\omega_k(t)$ and
$\sqrt{\left|\ddot{\omega}_k(t)\right| / \omega_k(t)}$
to be of the same order, this eventually yields
\begin{equation}
\frac{\left|\dot{\omega}_k(t)\right|}{\omega_k(t)} 
\ll \omega_k(t) \, .
\label{eq:adiab_cond_gen}
\end{equation}
Inequality~\eqref{eq:adiab_cond_gen} states that the rate of variation
of the instantaneous oscillation frequency has to be much smaller than
the frequency itself. In other words, $\omega_k(t)$ has to vary slowly
over an oscillation period, which is in agreement with the naive
discussion above. The adiabaticity condition~\eqref{eq:adiab_cond_gen}
actually depends on $k$. In order to understand in which range of
values of the momentum it is fulfilled, we first rewrite the left-hand
side as $\left|\dot{\omega}_k(t)\right| / \omega_k(t) =
\Omega_k \left|\dot{g}(t)\right| \rho / \hbar \omega_k^2(t)$.
Then, we recall that Eq.~\eqref{eq:adiab_cond_gen} has to be satisfied
at all times $t \geq t_0$ to maintain the adiabaticity for the whole
time evolution. After a bit of algebra we get
\begin{equation}
\frac{\Omega_k}{\sqrt{\dot{g}_{\mathrm{max}} \rho / \hbar}} \gg 
\left(1 + \frac{2 g_{\mathrm{min}} \rho}{\hbar \Omega_k} \right)^{-3/2}
\, ,
\label{eq:adiab_cond_highk_a}
\end{equation}
where $g_{\mathrm{min}} = \min_{t \geq t_0} g(t)$ and
$\dot{g}_{\mathrm{max}} = \max_{t \geq t_0} \left|\dot{g}(t)\right|$.
The latter quantity can be approximated as $\dot{g}_{\mathrm{max}}
\approx \Delta g / \tau$, where $\tau$ is the time it takes for the
coupling constant to change from its initial to its final value,
and $\Delta g$ is the corresponding variation of $g$ (in magnitude).
Since the right-hand side of Eq.~\eqref{eq:adiab_cond_highk_a}
is always smaller or equal to $1$, we find that the adiabatic approach
is accurate for the high-$k$ modes satisfying
\begin{equation}
k \gg \left(\frac{4 m^2 \rho\:\! \Delta g}{\hbar^3 \tau}\right)^{1/4} \, .
\label{eq:adiab_cond_highk_b}
\end{equation}

The adiabatic evolution of the tail of the momentum distribution
entails that the contact parameter defined in Sec.~\ref{subsec:exp_values}
exactly coincides at any time with its instantaneous value.
The latter is given by
\begin{equation}
\mathcal{C}(t) = \xi(t)^{-4} \, .
\label{eq:adiab_contact_par}
\end{equation}
On the other hand, the adiabatic approach is inadequate to study
the properties of the low-$k$ modes. This implies that the adiabatic
prediction for the quantum depletion,
\begin{equation}
\Delta N(t) = \frac{V}{3 \pi^2\xi^3(t)} \, ,
\label{eq:adiab_tot_depl}
\end{equation}
is always approximate. It is expected to be accurate only if $g(t)$
varies slowly enough in time.

\subsection{Physics at large evolution time and scattering formalism}
\label{subsec:scatt_form}
\begin{figure}[htb]
\centering
\includegraphics[scale=1]{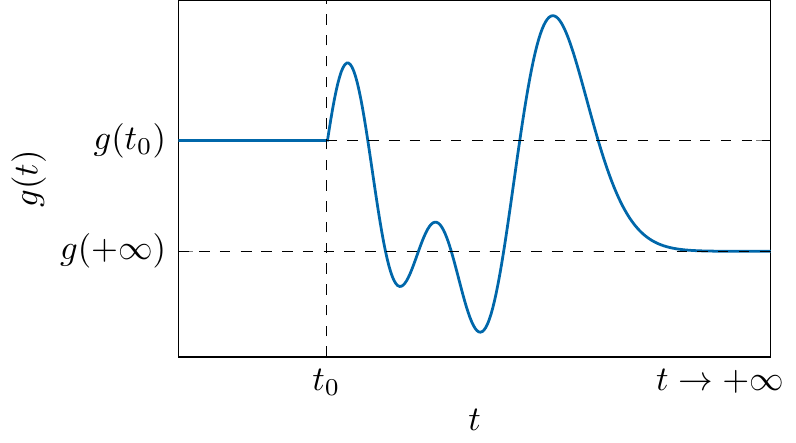}
\caption{Sketch of a typical $g(t)$ [with arbitrary time units]:
$g(t \to +\infty)$ is a constant and $g(t \leq t_0) = g(t_0)$.
$t_0 \to -\infty$ in the examples considered in
Sec.~\ref{sec:solvable_models}, in these cases one further
imposes that $g(t)$ tends to a constant value $g(-\infty)$ as
$t \to - \infty$, with $\dot{g}(-\infty) = 0$. In all the cases we
consider $g(t) \geq 0$ for all times.}
\label{fig:random_gt}
\end{figure}

One of the most interesting situations to analyze is that of a system whose
coupling coefficient tends to a constant value at long evolution times.
This problem can be directly mapped onto the quantum scattering of a particle
from a 1D potential barrier. In order to formulate this analogy in a
mathematically consistent way, it is necessary to define the time-dependent
coupling of our BEC over the whole time axis. Thus, for $t \leq t_0$ we take
$g$ to be constant and equal to $g(t_0)$. At $t=t_0$ the coupling starts
varying in time, and we denote by $g(t)$ its instantaneous value at $t > t_0$.
We further assume that $g(t)$ tends to a constant value $g(+\infty)$ as
$t \to + \infty$, with derivative $\dot{g}(+\infty) = 0$. This behavior is
shown in Fig.~\ref{fig:random_gt}.

The key observation is that, under the above assumptions, the TDHO
equation~\eqref{eq:quad_tdho_eq_gamma} has the same formal structure
as the Schr\"{o}dinger equation for a particle moving with zero energy
in a static external potential $V_\mathrm{ext}< 0$.  The analogy
relies on identifying, for the equivalent scattering problem, the physical
time $t$ with an effective position $x_\mathrm{eff} = t$ such that
$V_\mathrm{ext}(x_\mathrm{eff}) = - \omega_k^2(t)$. Here $\omega_k(t)$ is
given by Eq.~\eqref{eq:Bogo_freq}, with $g(t)$ behaving as illustrated in
Fig.~\ref{fig:random_gt}. Thus, we are effectively studying a 1D quantum
scattering problem along the time axis.

It is convenient to write the solution of Eq.~\eqref{eq:quad_tdho_eq_gamma}
in the form
\begin{equation}
\gamma_k(t,t_0) = e^{- i \omega_k(t_0) (t-t_0)} f_k(t,t_0) \, ,
\label{eq:scatt_gamma}
\end{equation}
where $f_k(t,t_0) = 1$ for $t \leq t_0$. Inserting the
ansatz~\eqref{eq:scatt_gamma} into Eq.~\eqref{eq:quad_tdho_eq_gamma}
one finds that $f_k$ obeys the second-order differential equation
\begin{equation}
\ddot{f}_k - 2i\omega_k(t_0) \dot{f}_k + [\omega_k^2(t) - 
\omega_k^2(t_0)] f_k = 0 \, .
\label{eq:scatt_fk_eq}
\end{equation}

At large times the Bogoliubov frequency~\eqref{eq:Bogo_freq}
approaches a constant value $\omega_k(+\infty)$. Consequently,
the asymptotic expression of the solution of
Eq.~\eqref{eq:quad_tdho_eq_gamma} takes the oscillating behavior
\begin{equation}
\begin{split}
\gamma_k(t,t_0) \underset{t \to + \infty}{=}
&{} e^{i \omega_k(t_0) t_0}
\sqrt{\frac{\omega_k(t_0)}{\omega_k(+\infty)}} \\
&{} \hspace{-1.4cm} \times \left[ \tl_k(t_0)\, e^{- i \omega_k(+\infty) t}
+ \tr_k(t_0)\, e^{i \omega_k(+\infty) t} \right] \, .
\end{split}
\label{eq:scatt_gamma_tpInf}
\end{equation}
Here $\tl_k(t_0)$ and $\tr_k(t_0)$ play the role, in the equivalent 1D
scattering problem, of the onward and backward transfer coefficient,
respectively (see, e.g., Ref.~\cite{LanLif03}).  The
relation between these coefficients can be derived from
Eq.~\eqref{eq:quad_tdho_curr_cons}, which is the analogous of the
current conservation. Because of the particular choice of the
prefactor in Eq.~\eqref{eq:scatt_gamma_tpInf}, at $t \to +\infty$
Eq.~\eqref{eq:quad_tdho_curr_cons} takes the simple and intuitive form
\begin{equation}
|\tl_k(t_0)|^2 - |\tr_k(t_0)|^2 = 1 \, .
\label{eq:scatt_tc_norm}
\end{equation}
The transfer coefficients generally depend on the specific functional
form of $g(t)$ and on the initial time $t_0$.
In Appendix~\ref{sec:change_initial_time} we show how, starting
from the knowledge of $\gamma_k(t,t_0)$, $\tl_k(t_0)$, and $\tr_k(t_0)$
for a given $t_0$, it is possible to calculate the same quantities for any
other initial time $t_0' > t_0$.

The asymptotic behavior of the time-propagated
weights~\eqref{eq:quad_uv_time} is easily deduced from the one
of $\gamma_k$ of Eq.~\eqref{eq:scatt_gamma_tpInf}:
\begin{subequations}
\label{eq:scatt_uv_time_tpInf}
\begin{align}
\begin{split}
U_k(t,t_0) \underset{t \to + \infty}{=} {}&{}
e^{i \omega_k(t_0) t_0}
\Big[ u_k(+\infty) \tl_k(t_0) e^{- i \omega_k(+\infty) t} \\
&{} \phantom{e^{i \omega_k(t_0) t_0} \Big[}
+ v_k(+\infty) \tr_k(t_0) e^{i \omega_k(+\infty) t} \Big] \, ,
\end{split}
\label{eq:scatt_u_time_tpInf} \\
\begin{split}
V_k(t,t_0) \underset{t \to + \infty}{=} {}&{}
e^{i \omega_k(t_0) t_0}
\Big[ v_k(+\infty) \tl_k(t_0) e^{- i \omega_k(+\infty) t} \\
&{} \phantom{e^{i \omega_k(t_0) t_0} \Big[}
+ u_k(+\infty) \tr_k(t_0) e^{i \omega_k(+\infty) t} \Big] \, .
\end{split}
\label{eq:scatt_v_time_tpInf}
\end{align}
\end{subequations}
In turn, these expressions enable one to directly
compute the asymptotic value of any observable. 

Let us consider a system initially in its ground state at zero
temperature. In this case the momentum distribution is
[from Eq.~\eqref{eq:depl_uv_time}]
\begin{equation}
\begin{split}
n_{\vec{k}}(t) \underset{t \to + \infty}{=}
{}&{} \Big| v_k(+\infty) \tl_k(t_0) e^{- i \omega_k(+\infty) t} \\
{}&{} \phantom{\Big|}
+ u_k(+\infty) \tr_k(t_0) e^{i \omega_k(+\infty) t} \Big|^2 \, ,
\end{split}
\label{eq:scatt_depl_T0_tpInf}
\end{equation}
the anti-diagonal part of the four-point correlation function
[from Eq.~\eqref{eq:dd_corr_uv_time}] reads as
\begin{equation}
\begin{split}
n^{(2)}_{\vec{k},-\vec{k}}(t) \underset{t \to + \infty}{=}
&{} \Big| u_k(+\infty) \tl_k(t_0) e^{-i \omega_k(+\infty) t} \\
& \phantom{{} \Big|}
+ v_k(+\infty) \tr_k(t_0) e^{i \omega_k(+\infty) t} \Big|^2 \\
&{} \times \Big| v_k(+\infty) \tl_k(t_0) e^{-i \omega_k(+\infty) t} \\
& \phantom{{} \times \Big|}
+ u_k(+\infty) \tr_k(t_0) e^{i \omega_k(+\infty) t} \Big|^2 \, ,
\end{split}
\label{eq:scatt_dd_corr_T0_tpInf}
\end{equation}
and the structure factor~\eqref{eq:str_fac_Bogo} is given by
\begin{equation}
\begin{split}
S(\vec{k},t) \underset{t \to + \infty}{=}
\frac{\Omega_k}{\omega_k(+\infty)} 
{}&{} \Big| \tl_k(t_0) e^{- i \omega_k(+\infty) t} \\
{}&{} \phantom{\Big|}
+ \tr_k(t_0) e^{i \omega_k(+\infty) t} \Big|^2 \, .
\end{split}
\label{eq:scatt_Sk_T0_tpInf}
\end{equation}
It is worth pointing out that, in general, $n_{\vec{k}}(t)$,
$n^{(2)}_{\vec{k},-\vec{k}}(t)$, and $S(\vec{k},t)$ keep
oscillating in time even at large $t$. As first discussed in
Ref.~\cite{Hun13} and re-analyzed below, these oscillations are
analogous to the cosmological Sakharov
oscillations~\cite{Sak65,Gri12b}, which originate in acoustic
vibrations in the primordial plasma of the early universe
before the epoch of recombination. Despite this time dependence, in 3D it is
possible to prove that the condensate depletion is time independent at
large $t$ (see Appendix~\ref{sec:asymp_depl}) and can be expressed as
\begin{equation}
\begin{split}
\Delta N(t) \underset{t \to + \infty}{=} V
\int \frac{d^3 k}{(2\pi)^3} 
\Big[ {}&{} |v_k(+\infty) \tl_k(t_0)|^2 \\
&{} + |u_k(+\infty) \tr_k(t_0)|^2 \Big] \, .
\end{split}
\label{eq:scatt_tot_depl_T0_tpInf}
\end{equation}
Concerning the Sakharov oscillations, the situation is particularly
simple when $g(+\infty) = 0$. In this case one has
$\omega_k(+\infty) = \Omega_k$, $u_k(+\infty) = 1$, and
$v_k(+\infty)=0$. Then, $n_{\vec{k}}$ and
$n^{(2)}_{\vec{k},-\vec{k}}$ become time independent at large time,
as clearly seen from Eqs.~\eqref{eq:scatt_depl_T0_tpInf}
and~\eqref{eq:scatt_dd_corr_T0_tpInf}:
\begin{subequations}
\label{eq:scatt_depl_and_dd_corr_T0_tpInf_gpInf0}
\begin{align}
n_{\vec{k}}(t) &{} \underset{t \to + \infty}{=} |\tr_k(t_0)|^2\, ,
\label{eq:scatt_depl_T0_tpInf_gpInf0}\\
n^{(2)}_{\vec{k},-\vec{k}}(t) &{} \underset{t \to + \infty}{=} 
|\tl_k(t_0)|^2 |\tr_k(t_0)|^2 \, .
\label{eq:scatt_dd_corr_T0_tpInf_gpInf0}
\end{align}
\end{subequations}
However, except for the exceptional cases discussed below,
$S(\vec{k},t)$ remains a time-dependent function, oscillating at period
$\pi/\omega_k(+\infty)$. This can be understood as resulting from the
creation of pairs of excitations during the epoch of time-dependent
$g(t)$. Because the system is homogeneous, momentum conservation
imposes that the excitations are created with opposite momenta
$\pm\vec{k}$. In our model these pairs, once created, survive when
$g(t)$ reaches its final zero value, and subsequently interfere
constructively each half of their common period.

A rather interesting situation occurs when one has total transmission
across the potential barrier for some $k$. This corresponds to the
condition $\tr_k(t_0) = 0$ or, equivalently, $|\tl_k(t_0)| = 1$.  For
these modes the asymptotic values of the time-propagated
weights~\eqref{eq:scatt_uv_time_tpInf} coincide, up to a phase factor,
with the corresponding instantaneous weights~\eqref{eq:Bogo_uv} at
$t \to + \infty$. As a consequence, for this specific values of $k$,
all the $k$-dependent observables [including the momentum
distribution~\eqref{eq:scatt_depl_T0_tpInf}, the four-point
correlation function~\eqref{eq:scatt_dd_corr_T0_tpInf}, and even the
structure factor~\eqref{eq:scatt_Sk_T0_tpInf}] are stationary
even if $g(+\infty) \neq 0$. Moreover, their values coincide with
those obtained for a BEC with time-independent coupling equal to
$g(+\infty)$. This observation is also related to the discussion of
Sec.~\ref{subsec:adiab_high_mom} about adiabatic evolution. In fact,
by noting that $\int_{t_0}^t d t' \, \omega_k(t') \sim \omega_k(+\infty) t
+ \mathrm{const}.$ for $t \to + \infty$, one finds that the large-$t$
behavior of Eq.~\eqref{eq:adiab_uv_time} is of the
kind~\eqref{eq:scatt_uv_time_tpInf} with $|\tl_k(t_0)| = 1$ and
$\tr_k(t_0) = 0$. This means that adiabaticity implies total
transmission across the potential barrier.

Up to now we have always assumed the initial time $t_0$ to be finite.
It actually makes sense to consider cases where $t_0 \to - \infty$,
as we do in Sec.~\ref{sec:solvable_models}. The only additional
requirements are that $g(t)$ tends to a constant value $g(-\infty)$
at large negative times and $\dot{g}(-\infty) = 0$. Notice that the global
phase factor $\exp[i \omega_k(t_0) t_0]$, first introduced in
Eq.~\eqref{eq:scatt_gamma} and subsequently entering the
time-propagated weights~\eqref{eq:scatt_uv_time_tpInf}, is ill defined
if $t_0 \to -\infty$. However, this phase factor does not represent a
problem because it systematically cancels when computing any observable
[for example, it no longer appears in Eqs.~\eqref{eq:scatt_depl_T0_tpInf},
\eqref{eq:scatt_dd_corr_T0_tpInf}, and~\eqref{eq:scatt_Sk_T0_tpInf}].

Finally, it is worth stressing that all the large-time expressions of the present
section have been derived within the framework of Bogoliubov theory. The latter
neglects the interaction between quasiparticles, which is expected to lead to
relaxation in our quantum many-body system at times $t \gg \hbar / [g(+\infty) \rho]$
(see Ref.~\cite{VanR18}). The study of such effects goes beyond the scope of this
work, within which the $t \to + \infty$ limit means that $t$ is much larger than
the typical scale of time variation of $g(t)$, while remaining smaller than the
thermalization time.

\section{Exactly solvable models}
\label{sec:solvable_models}
In this section we discuss in detail three examples where the TDHO
equation~\eqref{eq:quad_tdho_eq_gamma} can be solved analytically.
These are the steplike (Sec.~\ref{subsec:sl_coupling}), the
Woods-Saxon (Sec.~\ref{subsec:ws_coupling}), and the modified
P\"{o}schl-Teller coupling (Sec.~\ref{subsec:pt_coupling}). Other
solvable models may be considered, for instance the linear piecewise
$g(t)$ studied in Refs.~\cite{Ber14,Sch18}.

\subsection{Steplike coupling}
\label{subsec:sl_coupling}
The simplest case in which one can calculate everything analytically
is when the coupling constant has a steplike behavior. Let us
take\footnote{In the line of the discussion in Sec.~\ref{subsec:part_rep},
we note here that in this work we use the steplike
coupling~\eqref{eq:sl_gt} to approximately describe situations where
$\tau_{2\mathrm{B}} \ll \tau \ll \hbar / (g_{\mathrm{max}} \rho)$,
with $g_{\mathrm{max}} = \max_{t \geq t_0} g(t)$. The results obtained
in this way are in good agreement with experimental
observations~\cite{Hun13,Sch18}.}
\begin{equation}
g(t) =
\begin{cases}
g_0 & \mbox{if}\;\; t<0 \, , \\
g_1 & \mbox{if}\;\; t>0 \, .
\end{cases}
\label{eq:sl_gt}
\end{equation}
We indicate by $\omega_{k,0}$ and $\omega_{k,1}$ the Bogoliubov
frequency~\eqref{eq:Bogo_freq} before and after the jump, respectively;
the corresponding instantaneous weights~\eqref{eq:Bogo_uv} are denoted
by $u_{k,0}$, $v_{k,0}$ and $u_{k,1}$, $v_{k,1}$.

The problem is trivial for $t_0 > 0$, hence, in this section
we take $t_0 < 0$. For negative $t$, before the jump, 
the solution of Eq.~\eqref{eq:quad_tdho_eq_gamma} with
initial value~\eqref{eq:quad_tdho_eq_ic} is simply
$\gamma_k(t,t_0) = \exp[- i \omega_{k,0}(t-t_0)]$.
After the jump $\gamma_k$ must be a linear combination
of the two oscillating exponentials
$\exp(\pm i \omega_{k,1} t)$. By requiring the
continuity of $\gamma_k$ and its first-order derivative
one finds, for $t > 0$,
\begin{equation}
\begin{split}
\gamma_k(t,t_0) =
e^{i \omega_{k,0} t_0}
\sqrt{\frac{\omega_{k,0}}{\omega_{k,1}}}
\left( \tl_k \, e^{- i \omega_{k,1} t}
+ \tr_k \, e^{i \omega_{k,1} t} \right) \, .
\end{split}
\label{eq:sl_gamma}
\end{equation}
Here, the transfer coefficients are independent of $t_0$
and read as
\begin{equation}
\begin{split}
\tl_k &{} = \frac{1}{2} \left(
\sqrt{\frac{\omega_{k,1}}{\omega_{k,0}}}
+ \sqrt{\frac{\omega_{k,0}}{\omega_{k,1}}}
\right)  \, , \\
\tr_k &{} = \frac{1}{2} \left(
\sqrt{\frac{\omega_{k,1}}{\omega_{k,0}}}
- \sqrt{\frac{\omega_{k,0}}{\omega_{k,1}}}
\right)  \, .
\end{split}
\label{eq:sl_tc}
\end{equation}

The time-propagated weights~\eqref{eq:quad_uv_time} and all the
observables can be easily computed from the above formulas. In
particular, before the jump the observables are stationary.
Instead, for a system initially in its ground state at zero temperature, 
after the jump the momentum distribution, the anti-diagonal
four-point correlation function, and the structure factor are obtained by
inserting Eqs.~\eqref{eq:sl_tc} into~\eqref{eq:scatt_depl_T0_tpInf},
\eqref{eq:scatt_dd_corr_T0_tpInf},
and~\eqref{eq:scatt_Sk_T0_tpInf} (notice that all the asymptotic
formulas given in Sec.~\ref{subsec:scatt_form} exactly hold at any
$t > 0$ for a steplike coupling). This yields
\begin{equation}
\begin{split}
n_{\vec{k}}(t) =
\left|v_{k,0}\right|^2
+ \frac{g_1(g_1-g_0)\rho^2 \sin^2 \omega_{k,1} t}{
(\hbar\Omega_k + 2 g_1 \rho)
\sqrt{\hbar\Omega_k(\hbar\Omega_k + 2 g_0 \rho)}} \, ,
\end{split}
\label{eq:sl_depl_T0_t0mInf_tpInf}
\end{equation}
\begin{equation}
n^{(2)}_{\vec{k},-\vec{k}}(t) =  n_{\vec{k}}(t) [n_{\vec{k}}(t)+1] \, ,
\label{eq:sl_dd_corr_T0_t0mInf_tpInf}
\end{equation}
and
\begin{equation}
S(\vec{k},t) = \frac{\Omega_k}{\omega_{k,0}}
\left(1 + \frac{\omega_{k,0}^2-\omega_{k,1}^2}{\omega_{k,1}^2} 
\sin^2\omega_{k,1}t\right) \, .
\label{eq:sl_str_fac_T0_t0mInf_tpInf}
\end{equation}

Concerning the quantum depletion, after integration the first term
on the right-hand side of Eq.~\eqref{eq:sl_depl_T0_t0mInf_tpInf} returns
the depletion~\eqref{eq:tot_depl_T0_t0} of the condensate before
the jump. The integral of the second term can be easily computed
in the large-$t$ limit. To this purpose one needs to replace
$\sin^2 \omega_{k,1} t$ with $\frac{1}{2}$ (see Appendix~\ref{sec:asymp_depl})
and to change the integration variable from $k$ to $\tilde{k} =
\sqrt{(\hbar\Omega_k + 2g_0 \rho) / (2|g_1-g_0|\rho)}$.
The final result is
\begin{equation}
\Delta N(t) \underset{t \to + \infty}{=}
\frac{V}{3\pi^2\xi_0^3} + \frac{V}{2\pi^2\xi_1^3} \, \Delta \tilde{N}
\, .
\label{eq:sl_tot_depl_T0_t0mInf_tpInf}
\end{equation}
Here $\xi_{(0,1)} = \hbar/\sqrt{m g_{(0,1)} \rho}$ are the initial
and final healing lengths, and
\begin{equation}
\Delta \tilde{N}=
\begin{cases}
+ \frac{\sqrt{g_1^{2} - g_0^{2}}}{g_1} 
\arccot \sqrt{\frac{g_0}{g_1-g_0}} & \mbox{if}\;\; g_1 > g_0 \, , \\
- \frac{\sqrt{g_0^{2} - g_1^{2}}}{g_1} 
\arccoth \sqrt{\frac{g_0}{g_0-g_1}} & \mbox{if}\;\; g_1 < g_0 \, .
\end{cases}
\label{eq:sl_delta_N_tilde}
\end{equation}

Two limiting cases deserve special attention. If $g_1 = 0$ one has
$\omega_{k,1} = \Omega_k$, $u_{k,1} = 1$, $v_{k,1} = 0$,
$\tl_k = u_{k,0}$, and $\tr_k = v_{k,0}$. Inserting this relations
into Eqs.~\eqref{eq:scatt_uv_time_tpInf} one finds that the
time-propagated weights at $t > 0$ coincide, up to an oscillating phase,
with the instantaneous ones before the jump. This implies
that the momentum distribution remains frozen to its initial
value of Eq.~\eqref{eq:depl_uv_time_T0_t0} even at $t > 0$,
\begin{equation}
n_{\vec{k}}(t) = \left|v_{k,0}\right|^2 \, ,
\label{eq:sl_depl_T0_t0mInf_tpInf_g10}
\end{equation}
as one can check directly from Eq.~\eqref{eq:sl_depl_T0_t0mInf_tpInf}.
The same behavior is also exhibited by the four-point correlation
function and the quantum depletion. It is worth pointing out that
this prediction is consistent with the results of the recent
experiment~\cite{Lop17}. In this reference the authors measured
the momentum distribution of a uniform BEC after turning off the
interaction and the trapping potential, showing that it retains
the same value as the prequench one.

If, instead, $g_0 = 0$, i.e., $\omega_{k,0} = \Omega_k$, the transfer
coefficients~\eqref{eq:sl_tc} simplify to $\tl_k = u_{k,1}$ and
$\tr_k = - v_{k,1}$. The momentum distribution at $t > 0$ then becomes
\begin{equation}
n_{\vec{k}}(t) = \left|2 u_{k,1} v_{k,1}\right|^2 \sin^2 (\omega_{k,1} t) \, .
\label{eq:sl_depl_T0_t0mInf_tpInf_g00}
\end{equation}
and the asymptotic value of the quantum depletion is $\Delta N(t \to + \infty)
= V / (4 \pi \xi_1^3)$. This is larger by a factor $3\pi/4 \simeq 2.36$
than the depletion of a static condensate with coupling $g_1$.

\subsection{Woods-Saxon coupling}
\label{subsec:ws_coupling}
\begin{figure}[htb]
\centering
\includegraphics[scale=1]{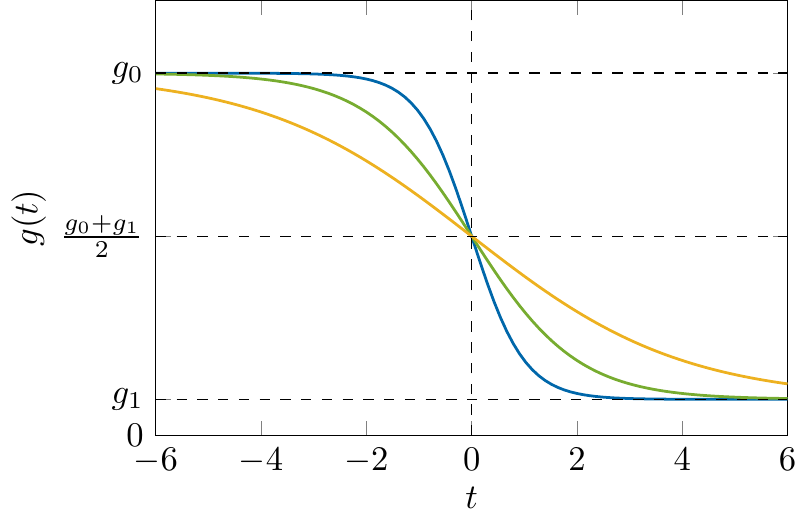}
\caption{Woods-Saxon coupling~\eqref{eq:ws_gt} as a function of time.
Here $g_1/g_0=0.1$ and $\tau = 0.5$ [blue (dark gray) curve],
$1.0$ [green (intermediate gray) curve], $2.0$ [yellow (light gray) curve].
Times are in units of $\hbar/(g_0 \rho)$. Although the drawing illustrates
the case $g_1<g_0$, the results of the present section hold true also
when $g_1>g_0$.}
\label{fig:ws_gt}
\end{figure}

Let us now consider a time-dependent coupling constant of the kind
\begin{equation}
g(t) = g_1 + \frac{g_0 - g_1}{1+e^{t/\tau}} \, ,
\label{eq:ws_gt}
\end{equation}
which has the same analytic form as the Woods-Saxon potential commonly
employed in nuclear physics. This coupling has been studied
numerically in Ref.~\cite{Rob17a}; the corresponding scattering problem
is known to be exactly solvable, cf.~\cite[\S 25, Problem 3]{LanLif03}.
The Woods-Saxon coupling varies smoothly
and monotonically from $g(-\infty) = g_0$ to $g(+\infty) = g_1$ (see
Fig.~\ref{fig:ws_gt}), the time scale for the change being fixed by
$\tau$. In the $\tau \to 0$ limit Eq.~\eqref{eq:ws_gt} tends to the
steplike coupling~\eqref{eq:sl_gt}, and all the formulas that we are
going to deduce in the present section reduce to the corresponding
ones of Sec.~\ref{subsec:sl_coupling}.  It is worth pointing out that,
although here we only address the $t_0 \to -\infty$ case, this choice
is not too restrictive. Indeed, once the solution for this special
case is known, one can use the procedure of
Appendix~\ref{sec:change_initial_time} to extend the results to
arbitrary $t_0$.

Equation~\eqref{eq:scatt_fk_eq} for the Woods-Saxon coupling
with $t_0 \to -\infty$ becomes
\begin{equation}
\ddot{f}_k - 2i\omega_{k,0} \dot{f}_k
+ (\omega_{k,1}^2 - \omega_{k,0}^2)
\frac{e^{t/\tau}}{1+e^{t/\tau}} f_k = 0 \, ,
\label{eq:ws_fk_eq}
\end{equation}
where we have adopted the same abbreviated notation $\omega_{k,0}
= \omega_k(-\infty)$ and $\omega_{k,1} = \omega_k(+\infty)$ as
in Sec.~\ref{subsec:sl_coupling}. After changing variable from
$t$ to $\zeta=-\exp(t/\tau)$ we obtain
\begin{equation}
\zeta(1-\zeta)\frac{d^2 f_k}{d \zeta^2} + 
[c_k-(1+a_k+b_k)\zeta] \frac{d f_k}{d \zeta} - a_k b_k f_k = 0 \, ,
\label{eq:ws_fk_zeta_eq}
\end{equation}
where we have defined the parameters
\begin{equation}
\begin{split}
a_k &{} = i(\omega_{k,1} - \omega_{k,0})\tau \, , \\
b_k &{} = -i(\omega_{k,1} + \omega_{k,0})\tau \, , \\
c_k &{} = 1-2i\omega_{k,0}\tau \, .
\end{split}
\label{eq:ws_abc}
\end{equation}
Equation~\eqref{eq:ws_fk_zeta_eq} corresponds to the well-known
hypergeometric differential equation~\cite{Abr65}. There are
two independent exact solutions available for this equation.
The first one is $f_k(\zeta) = {}_{2}F_1(a_k,b_k,c_k;\zeta)$,
where ${}_{2}F_1$ denotes the hypergeometric function. From
Eq.~\eqref{eq:scatt_gamma} one deduces the corresponding solution
of the TDHO equation~\eqref{eq:quad_tdho_eq_gamma},
\begin{equation}
\gamma_k(t,t_0 \to - \infty) = 
e^{-i \omega_{k,0} (t-t_0)} {}_{2}F_1(a_k,b_k,c_k;-e^{t/\tau}) \, .
\label{eq:ws_gamma}
\end{equation}
This expression fulfills the initial conditions~\eqref{eq:quad_tdho_eq_ic},
as can be easily checked by recalling that ${}_{2}F_1(a_k,b_k,c_k;0) = 1$.
Notice that the calculation of the derivative of $\gamma_k$ requires
the use of the relation $\frac{d}{d\zeta} \, {}_{2}F_1(a_k,b_k,c_k;\zeta)
= \frac{a_k b_k}{c_k} \, {}_{2}F_1(a_k+1,b_k+1,c_k+1;\zeta)$.

For completeness, we mention that the second independent solution of
Eq.~\eqref{eq:ws_fk_zeta_eq} is
$f_k(\zeta) = \zeta^{1-c_k}
{}_{2}F_1(1+a_k-c_k,1+b_k-c_k,2-c_k;\zeta)$.
Inserting this expression into Eq.~\eqref{eq:scatt_gamma} one gets (up
to an irrelevant constant factor) the complex conjugate of
Eq.~\eqref{eq:ws_gamma}. To verify this, one can first note that from
Eqs.~\eqref{eq:ws_abc} the three relations $1+a_k-c_k = b_k^*$,
$1+b_k-c_k = a_k^*$, and $2-c_k = c_k^*$ follow. The above statement
is then readily proved using the identity
${}_{2}F_1(b_k^*,a_k^*,c_k^*;\zeta) =
[{}_{2}F_1(a_k,b_k,c_k;\zeta)]^*$,
holding for real $\zeta$.  However this solution is not acceptable
because it does not fulfill the initial conditions~\eqref{eq:quad_tdho_eq_ic}.

Strictly speaking, the hypergeometric function
${}_{2}F_1(a_k,b_k,c_k;\zeta)$ is defined only for $\left|\zeta\right| < 1$.
This means that Eq.~\eqref{eq:ws_gamma} is valid only for negative $t$.
However, it can be extended by analytic continuation to $t \geq 0$,
as discussed in Appendix~\ref{sec:hyperg_func}. In particular, employing
the transformation~\eqref{eq:hyperg_mInf_1} one can see that the large-$t$
behavior is of the kind~\eqref{eq:scatt_gamma_tpInf}, with
\begin{equation}
\begin{split}
\tl_k (t_0 \to -\infty) &{} = 
\sqrt{\frac{b_k-a_k}{b_k+a_k}} 
\frac{\Gamma(c_k) \Gamma(b_k-a_k)}{\Gamma(b_k) \Gamma(c_k-a_k)} \, , \\
\tr_k (t_0 \to -\infty) &{} = \sqrt{\frac{b_k-a_k}{b_k+a_k}} 
\frac{\Gamma(c_k) \Gamma(a_k-b_k)}{\Gamma(a_k) \Gamma(c_k-b_k)} \, ,
\end{split}
\label{eq:ws_tc}
\end{equation}
where $\Gamma$ is the gamma function.

\begin{figure}[htb]
\centering
\includegraphics[scale=1]{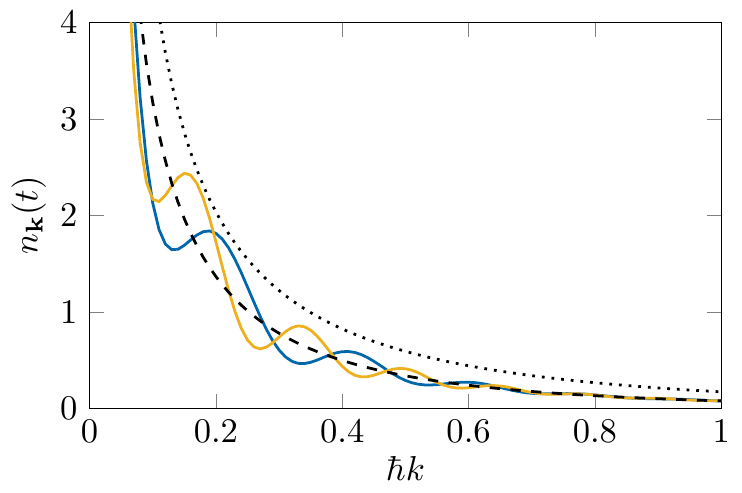}
\caption{Asymptotic value of the momentum distribution for the
Woods-Saxon coupling as a function of momentum. Here
$g_1/g_0 = 0.5$, $\tau = 1.0$, $t_0 \to - \infty$, and $t = 20.0$
[blue (dark gray) solid curve], $25.0$ [yellow (light gray) solid curve].
At each $\vec{k}$, $n_{\vec{k}}(t)$ oscillates around the value
$|v_k(+\infty) \tl_k(t_0)|^2 + |u_k(+\infty) \tr_k(t_0)|^2$,
that is indicated by the black dashed curve. For comparison, we also
plot the initial momentum distribution~\eqref{eq:depl_uv_time_T0_t0}
(black dotted curve). Times are in units of $\hbar / (g_0 \rho)$.
Momentum is in units of $\sqrt{m g_0 \rho}$.}
\label{fig:ws_mom_dist_tpInf_t}
\end{figure}

Now we have everything we need to compute exactly all the observables
of interest. We start by looking at the zero-temperature momentum
distribution at long evolution times. In Fig.~\ref{fig:ws_mom_dist_tpInf_t}
we plot the typical behavior of this quantity at two different (and large)
values of $t$. It can be clearly seen that it is non-monotonous and it varies
over time. We have checked that the results obtained from the exact
expression~\eqref{eq:depl_uv_time_T0} are in excellent agreement with
the asymptotic estimate~\eqref{eq:scatt_depl_T0_tpInf} in this large-$t$
regime.

\begin{figure}[htb]
\centering
\includegraphics[scale=1]{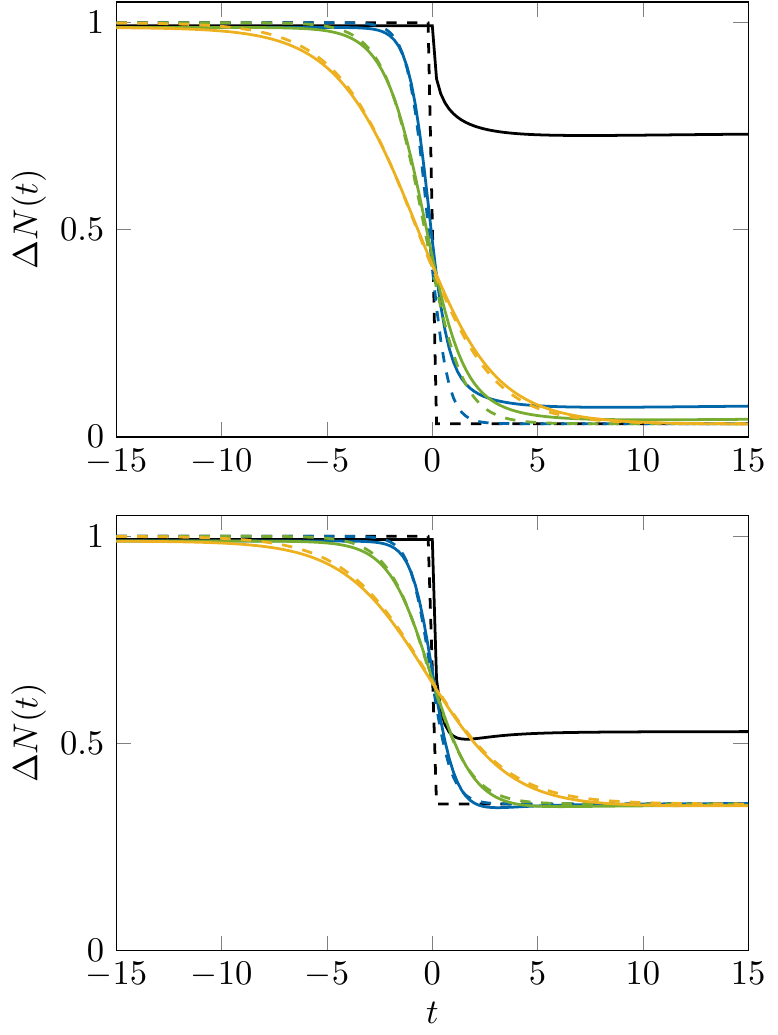}
\caption{Quantum depletion for the Woods-Saxon coupling
as a function of time for $g_1/g_0 = 0.1$ (top) and $g_1/g_0 = 0.5$
(bottom). Here $t_0 \to - \infty$ and $\tau = 0.0$ (black curves), $0.5$
[blue (dark gray) curves], $1.0$ [green (intermediate gray) curves],
$2.0$ [yellow (light gray) curves]. The solid and dashed lines show
the exact results and the adiabatic prediction~\eqref{eq:adiab_tot_depl},
respectively. Times are in units of $\hbar/(g_0 \rho)$. The depletion
is in units of $V / (3 \pi^2 \xi_0^3)$.}
\label{fig:ws_depl_t}
\end{figure}

Figure~\ref{fig:ws_depl_t} shows the quantum depletion~\eqref{eq:tot_depl}
as a function of time for two different choices of the final coupling strength
$g_1$. We consider several values of the characteristic time $\tau$, including
$\tau = 0$, which corresponds to the steplike coupling investigated in
Sec.~\ref{subsec:sl_coupling}. The exact results (solid lines) are compared
with the adiabatic prediction~\eqref{eq:adiab_tot_depl} (dashed lines). Notice
that the discrepancy between the two is significant when $\tau$ is small, and
particularly for $\tau = 0$; however, the agreement becomes extremely good
for the largest values of $\tau$ that we consider.

\begin{figure}[htb]
\centering
\includegraphics[scale=1]{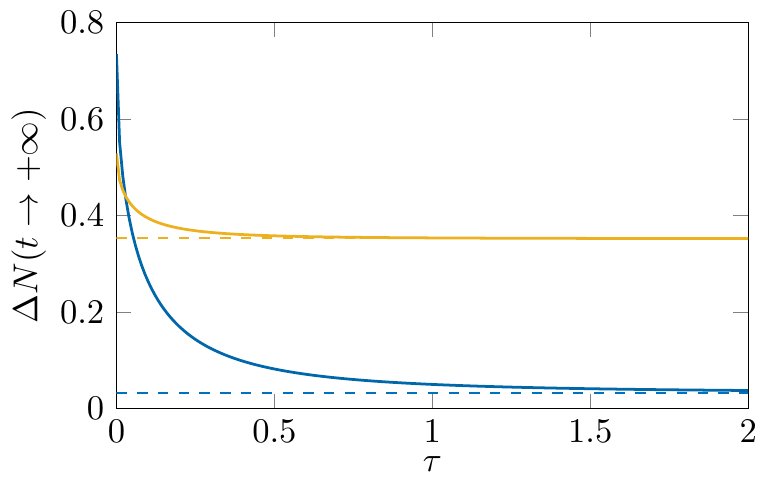}
\caption{Asymptotic value of the quantum depletion for the Woods-Saxon
coupling as a function of characteristic time. Here $t_0 \to - \infty$
and $g_1/g_0 = 0.1$ [blue (dark gray) solid curve], $0.5$ [yellow (light
gray) solid curve]. The dashed lines indicate the adiabatic prediction
$V / (3 \pi^2 \xi_1^3)$ corresponding to the two above choices of $g_1/g_0$.
Times are in units of $\hbar/(g_0 \rho)$. The depletion is in units
of $V / (3 \pi^2 \xi_0^3) = \Delta N(t\to-\infty)$.}
\label{fig:ws_depl_tpInf_tau}
\end{figure}

The existence of a crossover between nonadiabatic and adiabatic behavior 
of the quantum depletion as $\tau$ increases becomes more evident by
looking at Fig.~\ref{fig:ws_depl_tpInf_tau}. Here we plot the asymptotic
value~\eqref{eq:scatt_tot_depl_T0_tpInf} of the depletion as a function
of $\tau$. One can see that for $\tau \to 0$ it approaches the steplike
result~\eqref{eq:sl_tot_depl_T0_t0mInf_tpInf}, while as $\tau \to + \infty$
it goes asymptotically to the adiabatic value $V / (3 \pi^2 \xi_1^3)$.

Let us now consider the simplest case $g_1 = 0$ and a system initially
in its ground state at $T=0$. According to
Eq.~\eqref{eq:scatt_depl_T0_tpInf_gpInf0}, the stationary value of the
$T=0$ momentum distribution at large $t$ coincides with the square
modulus of the coefficient $\tr_k(t_0 \to -\infty)$ given in
Eq.~\eqref{eq:ws_tc}. The analytic formula for this quantity can be
significantly simplified using the identities
$[\Gamma(z)]^* = \Gamma(z^*)$, $\Gamma(1+z) = z \Gamma(z)$,
$\Gamma(z) \Gamma(1-z) = \pi / \sin(\pi z)$ (reflection formula), and
$\sin(i z) = i \sinh z$. Taking the expressions~\eqref{eq:ws_abc} of
the parameters $a_k$, $b_k$, and $c_k$ into account and setting
$\omega_{k,1} = \Omega_k$, one eventually obtains
\begin{equation}
n_{\vec{k}}(t) \underset{t \to + \infty}{=}
\frac{\sinh^2[\pi(\Omega_k - \omega_{k,0})\tau]}
{\sinh(2\pi\Omega_k\tau) \sinh(2\pi\omega_{k,0}\tau)} \, .
\label{eq:ws_depl_t0mInf_tpInf_gInf0}
\end{equation}
Notice that in the $k \to 0$ regime the behavior of the asymptotic
momentum distribution, $n_{\vec{k}}(t \to + \infty) \sim m c_0
/ 2 \hbar k$ [here $c_0 = c(-\infty)$], is the same as at the
initial time $t_0 \to -\infty$. This is in full agreement with
the general findings of Sec.~\ref{subsec:freezing_low_mom}.
In the opposite limit $k \to + \infty$ one has instead
$n_{\vec{k}}(t \to + \infty) \sim 4 \sinh^2(\pi m c_0^2 \tau / \hbar)
\exp[-2\pi (\hbar \tau / m) k^2]$.

The anti-diagonal four-point correlation function for $g_1 = 0$
can be calculated starting from Eq.~\eqref{eq:scatt_dd_corr_T0_tpInf_gpInf0}.
Proceeding as we did for the momentum distribution, we end up with
\begin{equation}
\begin{split}
n^{(2)}_{\vec{k},-\vec{k}}(t) \underset{t \to + \infty}{=} {}&{} \\
{}&{} \hspace{-1.5cm} \left\{ \frac{\sinh[\pi(\Omega_k - \omega_{k,0})\tau]
\sinh[\pi(\Omega_k + \omega_{k,0})\tau] }
{\sinh(2\pi\Omega_k\tau) \sinh(2\pi\omega_{k,0}\tau)} \right\}^2
\, .
\end{split}
\label{eq:ws_dd_corr_t0mInf_tpInf_gInf0}
\end{equation}

\subsection{Modified P\"{o}schl-Teller coupling}
\label{subsec:pt_coupling}
\begin{figure}[htb]
\centering
\includegraphics[scale=1]{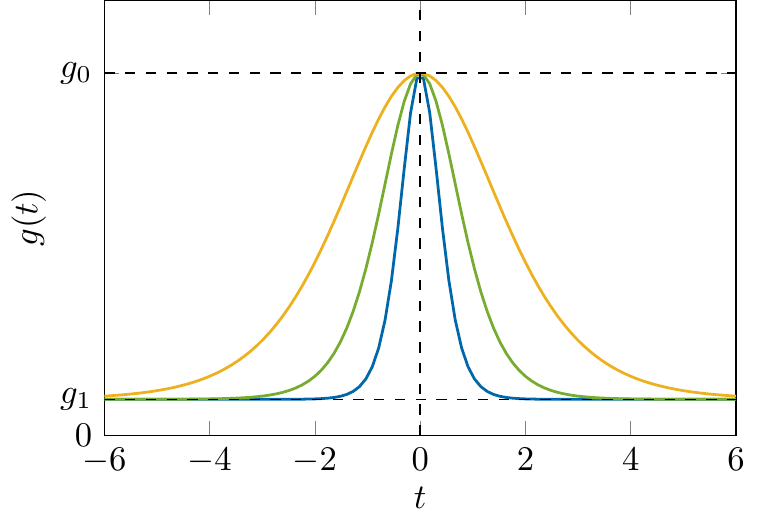}
\caption{Modified P\"{o}schl-Teller coupling~\eqref{eq:pt_gt} as a function
of time. Here $g_1/g_0=0.1$ and $\tau = 0.5$ [blue (dark gray) curve],
$1.0$ [green (intermediate gray) curve], $2.0$ [yellow (light gray) curve].
Times are in units of $\hbar/(g_0 \rho)$.}
\label{fig:pt_gt}
\end{figure}

The modified P\"{o}schl-Teller coupling is defined as
\begin{equation}
g(t) = g_1 + \frac{g_0-g_1}{\cosh^2(t/\tau)} \, .
\label{eq:pt_gt}
\end{equation}
This coupling changes monotonically starting from the value $g_1$
at $t \to - \infty$, reaches the value $g_0$ at $t=0$, and then goes back
to $g_1$ for $t \to + \infty$. $g(t)$ is an even function of time, and attains
a minimum (maximum) at $t=0$ when $g_0 < g_1$ ($g_0 > g_1$ as illustrated
in Fig.~\ref{fig:pt_gt}). As for the Woods-Saxon coupling, a finite scale $\tau$
quantifies how rapidly the coupling changes over time.

Equation~\eqref{eq:scatt_fk_eq} with the modified P\"{o}schl-Teller
coupling and $t_0 \to - \infty$ reads
\begin{equation}
\ddot{f}_k - 2i\omega_{k,1} \dot{f}_k 
+ \frac{\omega_{k,0}^2 - \omega_{k,1}^2}{\cosh^{2}(t/\tau)} f_k = 0 \, .
\label{eq:pt_fk_eq}
\end{equation}
Notice that here and in the rest of the present section we are using
the notation $\omega_{k,0} = \omega_k(0)$ and $\omega_{k,1} =
\omega_k(\pm\infty)$. We deal with Eq.~\eqref{eq:pt_fk_eq} in a way
similar to the one illustrated in~\cite[\S 23, Problem 5 and \S 25,
Problem 4]{LanLif03}. Changing the variable to
$\zeta = [1 + \tanh(t/\tau)]/2$ it becomes
\begin{equation}
\zeta(1-\zeta)\frac{d^2 f_k}{d \zeta^2} + 
(c_k - 2\zeta) \frac{d f_k}{d \zeta} + s_k(s_k+1) f_k = 0 \, .
\label{eq:pt_fk_zeta_eq}
\end{equation}
Here we have introduced the two quantities
\begin{equation}
\begin{split}
& s_k = \frac{1}{2} 
\left[\sqrt{4(\omega_{k,0}^2 - \omega_{k,1}^2)\tau^2 + 1}
- 1\right] \, ,\\
& c_k = 1 - i\omega_{k,1}\tau \, .
\end{split}
\label{eq:pt_sc}
\end{equation}
Notice that $s_k$ is always a real positive number if $g_0 > g_1$.
It is instead real and negative if $g_1 > g_0$ and
$4(\omega_{k,0}^2 - \omega_{k,1}^2) \tau^2 + 1 \geq 0$,
that is, $k \leq \sqrt{m / [4 (g_1-g_0) \rho \tau^2]}$.
In all the other cases $s_k$ becomes complex.

Equation~\eqref{eq:pt_fk_zeta_eq} has the same form as the
hypergeometric equation~\eqref{eq:ws_fk_zeta_eq} with $a_k = -s_k$
and $b_k = s_k + 1$. Consequently, its solutions are expressed
in terms of hypergeometric functions. The first independent solution
that we consider is $f_k(\zeta) = {}_{2}F_1(-s_k,s_k+1,c_k;\zeta)$.
By plugging it into Eq.~\eqref{eq:scatt_gamma} one gets
\begin{equation}
\begin{split}
\gamma_k(t,t_0 \to - \infty) = {}&{}
e^{-i \omega_{k,1} (t-t_0)} \\
&{} \times
{}_{2}F_1\left(-s_k,s_k+1,c_k;
\textstyle\frac{e^{2t/\tau}}{e^{2t/\tau}+1}\right) \, .
\end{split}
\label{eq:pt_gamma}
\end{equation}
This function satisfies both the TDHO
equation~\eqref{eq:quad_tdho_eq_gamma} and the initial
conditions~\eqref{eq:quad_tdho_eq_ic}. Thus, it will be used
in all the calculations of the remaining part of the present
section.

The second independent solution of Eq.~\eqref{eq:pt_fk_zeta_eq} is
$f_k(\zeta) = \zeta^{1-c_k}
{}_{2}F_1(1-s_k-c_k,2+s_k-c_k,2-c_k;\zeta)$.
Here, the same thing happens as for the Woods-Saxon coupling: this
second solution is not acceptable because it does not fulfill the
initial conditions~\eqref{eq:quad_tdho_eq_ic}.\footnote{Inserting
the solution $f_k(\zeta) = \zeta^{1-c_k}
{}_{2}F_1(1-s_k-c_k,2+s_k-c_k,2-c_k;\zeta)$ into
Eq.~\eqref{eq:scatt_gamma} one obtains the complex conjugate of
Eq.~\eqref{eq:pt_gamma}. In order to check this, it is first
convenient to use the Euler
transformation~\eqref{eq:hyperg_euler_trans} to rewrite the above
expression as $f_k(\zeta) = [\zeta/(1-\zeta)]^{1-c_k}
{}_{2}F_1(s_k+1,-s_k,2-c_k;\zeta)$. Then, the proof follows from
the identities $[\zeta/(1-\zeta)]^{1-c_k} = e^{2 i \omega_{k,1} t}$,
$s_k=s_k^*$ (for real $s_k$) or $s_k+1 = -s_k^*$ (for complex $s_k$),
$2-c_k = c_k^*$, and ${}_{2}F_1(-s_k^*,s_k^*+1,c_k^*;\zeta) =
[{}_{2}F_1(-s_k,s_k+1,c_k;\zeta)]^*$ (for real $\zeta$).}

\begin{figure}[htb]
\centering
\includegraphics[scale=1]{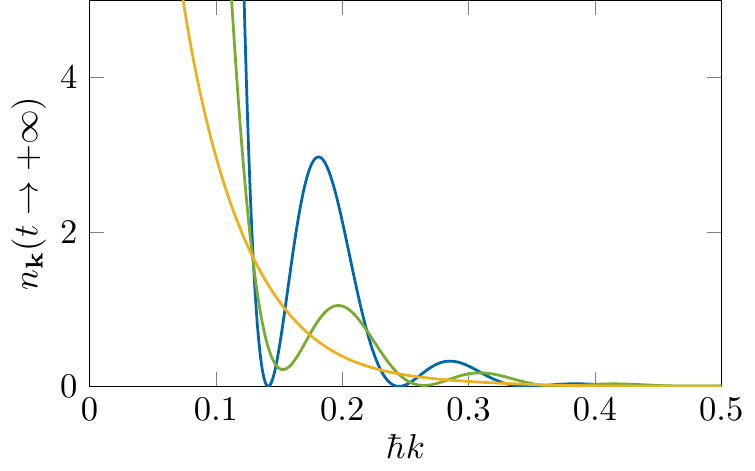}
\caption{Asymptotic value of the momentum distribution for the modified
P\"{o}schl-Teller coupling as a function of momentum. Here, $g_1=0$,
$\tau = 10.0$, $t_0 = - \infty$ [blue (dark gray) curve], $-20.0$
[green (intermediate gray) curve], $0.0$ [yellow (light gray) curve].
Times are in units of $\hbar/(g_0 \rho)$. Momentum is in units of
$\sqrt{m g_0 \rho}$.}
\label{fig:pt_mom_dist_tpInf_t0}
\end{figure}

The transfer coefficients for the modified P\"{o}schl-Teller coupling
can be obtained by applying the transformation~\eqref{eq:hyperg_mInf_2}
to Eq.~\eqref{eq:pt_gamma} and taking the large-$t$ limit. The result is
\begin{equation}
\begin{split}
\tl_k(t_0 \to - \infty) &{} = 
\frac{\Gamma(c_k) \Gamma(c_k-1)}{\Gamma(c_k+s_k) \Gamma(c_k-s_k-1)} \, , \\
\tr_k(t_0 \to - \infty) &{} = 
\frac{\Gamma(c_k) \Gamma(1-c_k)}{\Gamma(-s_k) \Gamma(s_k+1)} \, .
\end{split}
\label{eq:pt_tc}
\end{equation}
An interesting consequence of Eq.~\eqref{eq:pt_tc} is that the modified
P\"{o}schl-Teller coupling supports total transmission if $g_0 > g_1$.
Indeed, $\Gamma(-s_k)$ diverges if $s_k$ is a non-negative integer,
which entails the vanishing of $\tr_k(t_0 \to - \infty)$. This happens
whenever
\begin{equation}
k = \sqrt{\frac{p(p+1) m}{(g_0-g_1)\rho \tau^2}} \, ,
\quad p \in \mathbb{N} \, .
\label{eq:pt_k_res}
\end{equation}
As discussed at the end of Sec.~\ref{subsec:scatt_form}, when
$k$ fulfills the resonance condition~\eqref{eq:pt_k_res} the
$k$-dependent observables are stationary at large times; their
values coincide with those of a BEC with constant coupling $g_1$.
Additionally, if $g_1 = 0$, the momentum distribution and the
four-point correlation function have a non-monotonous behavior
even if they are stationary at all $k$'s, and they vanish at the
resonant momenta~\eqref{eq:pt_k_res} [see
Eqs.~\eqref{eq:pt_depl_t0mInf_tpInf_gInf0}
and~\eqref{eq:pt_dd_corr_t0mInf_tpInf_gInf0} below]. All these
considerations hold for the $t_0 \to - \infty$ case, but the
scenario can partly persist if one switches to finite $t_0$
(see Appendix~\ref{sec:change_initial_time}). As shown in
Fig.~\ref{fig:pt_mom_dist_tpInf_t0}, resonant and quasi-resonant
situations are possible if $t_0$ is not too close to $0$ or positive.

\begin{figure}[htb]
\centering
\includegraphics[scale=1]{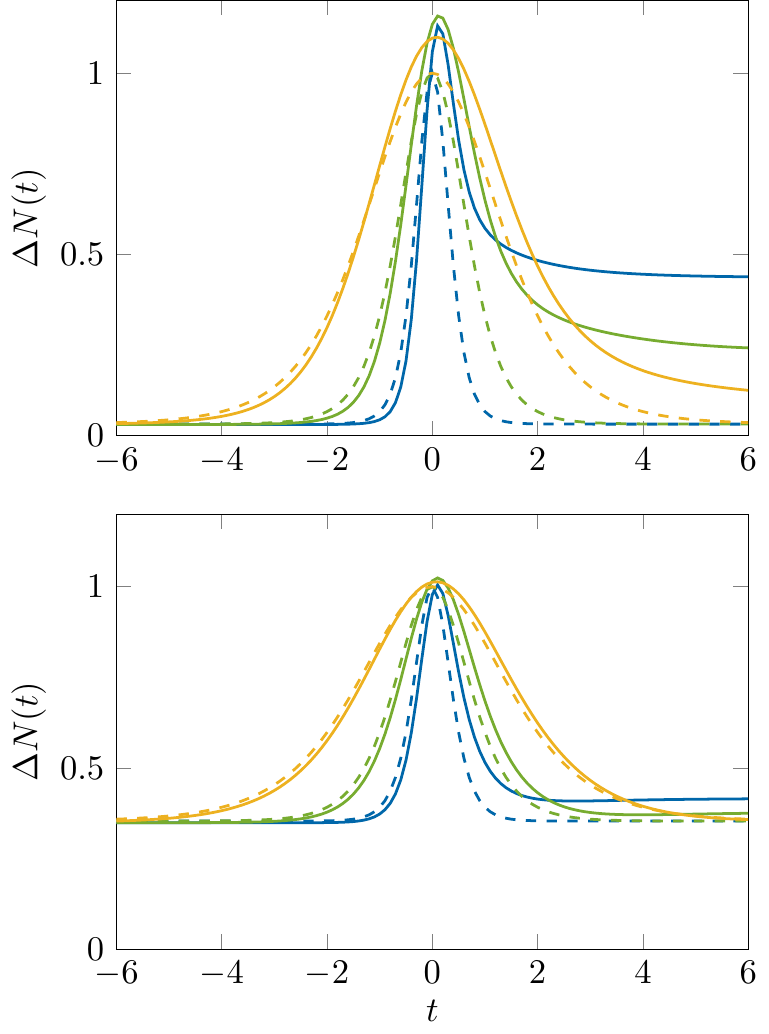}
\caption{Quantum depletion for the modified P\"{o}schl-Teller
coupling as a function of time for $g_1/g_0 = 0.1$ (top) and
$g_1/g_0 = 0.5$ (bottom). Here, $t_0 \to - \infty$ and $\tau = 0.5$
[blue (dark gray) curves], $1.0$ [green (intermediate gray) curves],
$2.0$ [yellow (light gray) curves]. The solid and dashed lines show
the exact results and the adiabatic prediction~\eqref{eq:adiab_tot_depl},
respectively. Times are in units of $\hbar/(g_0 \rho)$. The depletion
is in units of $V / (3 \pi^2 \xi_0^3)$.}
\label{fig:pt_depl_t}
\end{figure}

The behavior of the quantum depletion~\eqref{eq:tot_depl} as a
function of time is shown in Fig.~\ref{fig:pt_depl_t} for several
values of $\tau$ and $g_1/g_0$ (here, and in all the rest of this
section, we consider a system initially in its ground state at
$T=0$). As already seen in Sec.~\ref{subsec:ws_coupling} for the
Woods-Saxons coupling, the exact prediction gets closer and closer to
the adiabatic one~\eqref{eq:adiab_tot_depl} as $\tau$ increases.

\begin{figure}[htb]
\centering
\includegraphics[scale=1]{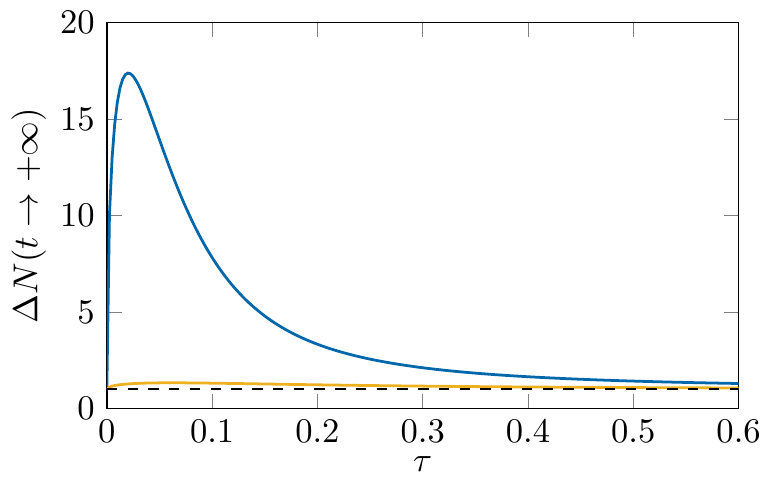}
\caption{Asymptotic value of the quantum depletion for the modified
P\"{o}schl-Teller coupling as a function of characteristic time.
Here, $t_0 \to - \infty$ and $g_1/g_0 = 0.1$ [blue (dark gray) solid curve],
$0.5$ [yellow (light gray) solid curve]. The black dashed line indicates
the adiabatic prediction $V / (3 \pi^2 \xi_1^3)$.
Times are in units of $\hbar/(g_1 \rho)$.
The depletion is in units of $V / (3 \pi^2 \xi_1^3) = \Delta N(t\to-\infty)$.}
\label{fig:pt_depl_tpInf_tau}
\end{figure}

In Fig.~\ref{fig:pt_depl_tpInf_tau} we plot the asymptotic
depletion~\eqref{eq:scatt_tot_depl_T0_tpInf} as a function of $\tau$.
The behavior of the curve follows from the property that the initial
and final values of the coupling coincide. For small $\tau$,
a large number of modes fall in the small-$k$ part of the momentum
distribution that stays frozen. Consequently, the depletion remains
approximately constant in time. At large $\tau$, instead, the majority
of the modes are in the large-$k$ tail that behaves adiabatically. Thus,
the depletion goes back to its initial value at the end of the time
evolution. For these two reasons, the values of the depletion before
and after time evolution can differ significantly from each other
only for intermediate $\tau$.

Let us now go back briefly to the case $t_0 = -\infty$ and $g_1 = 0$.
According to Eq.~\eqref{eq:scatt_depl_T0_tpInf_gpInf0}, the stationary
value of the momentum distribution at zero temperature can be found
by computing the square modulus of the coefficient
$\tr_k(t_0 \to - \infty)$ given in Eq.~\eqref{eq:pt_tc} and setting
$\omega_{k,1} = \Omega_k$. The calculation is very similar to the one
that led us to Eq.~\eqref{eq:ws_depl_t0mInf_tpInf_gInf0}. The result is
\begin{equation}
n_{\vec{k}}(t) \underset{t \to + \infty}{=}
\frac{\cos^2\left[ \frac{\pi}{2} \sqrt{4 (\omega_{k,0}^2 - 
\Omega_k^2)\tau^2 + 1} \, \right]}{\sinh^2 (\pi \Omega_k\tau)} \, .
\label{eq:pt_depl_t0mInf_tpInf_gInf0}
\end{equation}
At $k \to + \infty$, $n_{\vec{k}}(t \to + \infty)$ decreases
exponentially to $0$, very roughly as $\exp[- \pi (\hbar \tau / m) k^2]$.
Instead, at $k \to 0$ the momentum distribution tends to a finite value
$n_{\vec{k}=0}(t \to + \infty) = (2 g_0 \rho \tau / \hbar)^2$.
The latter could equally be obtained by calculating the
integral~\eqref{eq:pert_v_time_g00} and taking its square
modulus. The anti-diagonal four-point correlation function at zero
temperature can be deduced from
Eq.~\eqref{eq:scatt_dd_corr_T0_tpInf_gpInf0} and reads as
\begin{equation}
n^{(2)}_{\vec{k},-\vec{k}}(t) \underset{t \to + \infty}{=}
n_{\vec{k}}(t) [n_{\vec{k}}(t)+1] \, .
\label{eq:pt_dd_corr_t0mInf_tpInf_gInf0}
\end{equation}

\section{Quantum coherence in 3D and 1D}
\label{sec:coh_prop}
The degree of quantum coherence of the system is characterized by its
one-body density matrix, defined in terms of the field operator
$\hat{\Psi}(\vec{r},t)$ as~\cite{Dal99,PitStr16}
\begin{equation}
\rho^{(1)}(\vec{r},\vec{r}',t) =
\langle \hat{\Psi}^\dagger(\vec{r},t) \hat{\Psi}(\vec{r}',t) \rangle \, .
\label{eq:OBDMDef}
\end{equation}
In the weakly interacting regime considered in this work, $\rho^{(1)}$
can be computed in dimension $d=3$ within the standard Bogoliubov theory
of linearized quantum fluctuations. In lower dimension, when $d = 1$ or $2$,
the large phase fluctuations of $\hat{\Psi}(\vec{r},t)$ drastically affect the
phase coherence of the system and destroy Bose-Einstein condensation,
the existence of which is at the heart of standard Bogoliubov theory.
However, even in this case, generalized Bogoliubov
theories~\cite{Pop72,Pop83,Mor03,Lar13} may be employed for calculating
$\rho^{(1)}$ in the limit of weak interactions and small density fluctuations. 
The correct treatment, valid in any dimension and for any separation
$|\vec{r}-\vec{r}'|$, yields for our nonequilibrium system~\cite{Lar16,Lar18}
\begin{equation}
\rho^{(1)}(\vec{r},\vec{r}',t) =
\rho\,\exp\!\bigg[{-}\frac{\Delta\rho(\vec{r},\vec{r},t)
- \Delta\rho(\vec{r},\vec{r}',t)}{\rho}\bigg] \, ,
\label{eq:OBDM}
\end{equation}
where $\Delta\rho(\vec{r},\vec{r}',t)$ is expressed in terms of
the quench-dependent Bogoliubov momentum distribution
$n_{\vec{k}}(t)$ as
\begin{equation}
\Delta\rho(\vec{r},\vec{r}',t) =
\int\frac{d^{d}k}{(2\pi)^{d}} \, n_{\vec{k}}(t)
\cos[\vec{k}\cdot(\vec{r}-\vec{r}')] \, ,
\label{eq:GeneralizedDepletion}
\end{equation}
and we recall that $\rho = \rho^{(1)}(\vec{r},\vec{r},t)$ is the density of
the gas, here homogeneous.

When $d=3$, the argument of the exponential in Eq.~\eqref{eq:OBDM}
is small and one may approximate the one-body density matrix by
\begin{equation}
\rho^{(1)}(\vec{r},\vec{r}',t) \simeq
\rho - \Delta\rho(\vec{r},\vec{r},t) + \Delta\rho(\vec{r},\vec{r}',t) \, .
\label{eq:rho-DL}
\end{equation}
This expression easily compares with the Bogoliubov prediction for
$\rho^{(1)}$~\cite{Dal99,PitStr16}, only valid in 3D. In particular,
$\rho-\Delta\rho(\vec{r},\vec{r},t) = \rho_0(t)$
is nothing but the density of the condensate, obtained by subtracting to the
mean density $\rho$ of the gas the quantum depletion
\begin{equation}
\Delta\rho(\vec{r},\vec{r},t) =
\int \frac{d^{3}k}{(2\pi)^{3}} \, n_{\vec{k}}(t) = \frac{\Delta N(t)}{V}
\label{eq:Depletion}
\end{equation}
[see Eq.~\eqref{eq:tot_depl}]. In this case, off-diagonal long-range
order is achieved since the remaining term in Eq.~\eqref{eq:rho-DL},
$\Delta\rho(\vec{r},\vec{r}',t)$, vanishes for distant $\vec{r}$ and
$\vec{r}'$, and in this case $\rho^{(1)}$ tends to a finite value equal
to the density of the condensate:
\begin{equation}
\rho^{(1)}(\vec{r},\vec{r}',t)
\xrightarrow[|\vec{r}-\vec{r}'|\to + \infty]{} \rho_0(t) \, .
\end{equation}
Note that the long-time density $\rho_0(t \to +\infty)$ of the
condensate is well defined since, as shown in Sec.~\ref{subsec:scatt_form},
$\Delta N(t)$ tends to a constant as $t \to +\infty$ [see
Eq.~\eqref{eq:scatt_tot_depl_T0_tpInf}].

In lower dimension ($d=1$ or $2$), the large phase fluctuations of the
field operator rule out Bose-Einstein condensation
\cite{Mer66,Hoh67,Pit91,Fis02}, and the first-order
expansion~\eqref{eq:rho-DL} is not possible. In this case we must
rely on the exact expression~\eqref{eq:OBDM} to calculate the one-body
density matrix of the system.
In the following we focus on the 1D geometry where one has
\begin{equation}
\begin{split}
\rho^{(1)}(x,x',t) &{} =
\rho\,\exp\!\bigg\{{-}\frac{1}{\rho} \int \frac{dk}{2\pi} \, n_k(t) \\
& \phantom{{}=} \times [1-\cos(k |x-x'|)] \bigg\} \, .
\end{split}
\label{eq:OBDMQuench}
\end{equation}
Here, we substituted the vector notations $\vec{r}$, $\vec{r}'$,
and $\vec{k}$ with $x$, $x'$, and $k$, respectively. The momentum
distribution $n_k(t)$ appearing in the above expression depends on
the type of quench. For a system initially in a thermal state
$n_k(t)$ as given by Eq.~\eqref{eq:depl_uv_time}
may be separated in a zero-temperature term $n_k(t)|_0$ and a
remaining contribution $n_k(t)|_T$ that vanishes at zero temperature:
\begin{subequations}
\label{eq:GenericMomentumDistribution}
\begin{align}
n_k(t) &{} = n_k(t)|_0 + n_k(t)|_T \, , 
\label{eq:GenericMomentumDistribution-a} \\
n_k(t)|_0 &{} = |V_k(t,t_0)|^2 \, ,
\label{eq:GenericMomentumDistribution-b} \\
n_k(t)|_T &{} = \frac{1 + 2 |V_k(t,t_0)|^2}
{\exp[\hbar\omega_k(t_0)/(k_{\mathrm{B}}T)]-1} \, .
\label{eq:GenericMomentumDistribution-c}
\end{align}
\end{subequations}
In the next two sections, we separately consider the case of the
steplike coupling~\eqref{eq:sl_gt} (Sec.~\ref{subsec:1Dstep}) and of
the Woods-Saxon coupling~\eqref{eq:ws_gt}
(Sec.~\ref{subsec:1DWS}). Calculations are mostly performed at zero
temperature, i.e., when $n_k(t) = n_k(t)|_0$. A quantitative analysis
of how temperature affects $\rho^{(1)}$ is explicitly provided for
the steplike coupling in Sec.~\ref{subsec:1Dstep}.

\subsection{Steplike coupling in 1D}
\label{subsec:1Dstep}
This quench protocol was already widely investigated for 1D systems
that initially do not interact ($g_0 = 0$) and evolve after the quench
with a Lieb-Liniger--type Hamiltonian ($g_1 \neq 0$). In this case,
both the regimes of weak~\cite{Lar16,Lar18}, strong~\cite{Gri10,
Kor14,Foi17}, and generic~\cite{DeN14,Pir16} postquench coupling
were analyzed. Here, we focus on the case where $g_0$ is nonzero
and $g_1$ is arbitrary (and thus possibly zero). Such a model
is relevant for describing the nonequilibrium dynamics of an
ultracold gas of weakly interacting atom bosons aligned along the $x$
axis after some arbitrary change of the stiffness of the transverse
confining potential
$V_{\mathrm{trap}}(y,z,t) = m \omega_\perp^2\!(t)(y^2+z^2) / 2$.
In this case, the 3D $s$-wave scattering length $a$ does not depend on
time (as assumed in the discussion of Sec.~\ref{subsec:part_rep}), but
the effective 1D coupling constant $g(t)$ does, since it relates to the
transverse trapping frequency as
$g(t) = 2 \hbar \omega_\perp\!(t) a$~\cite{Ols98}.

\begin{figure}[t!]
\includegraphics[width=\linewidth]{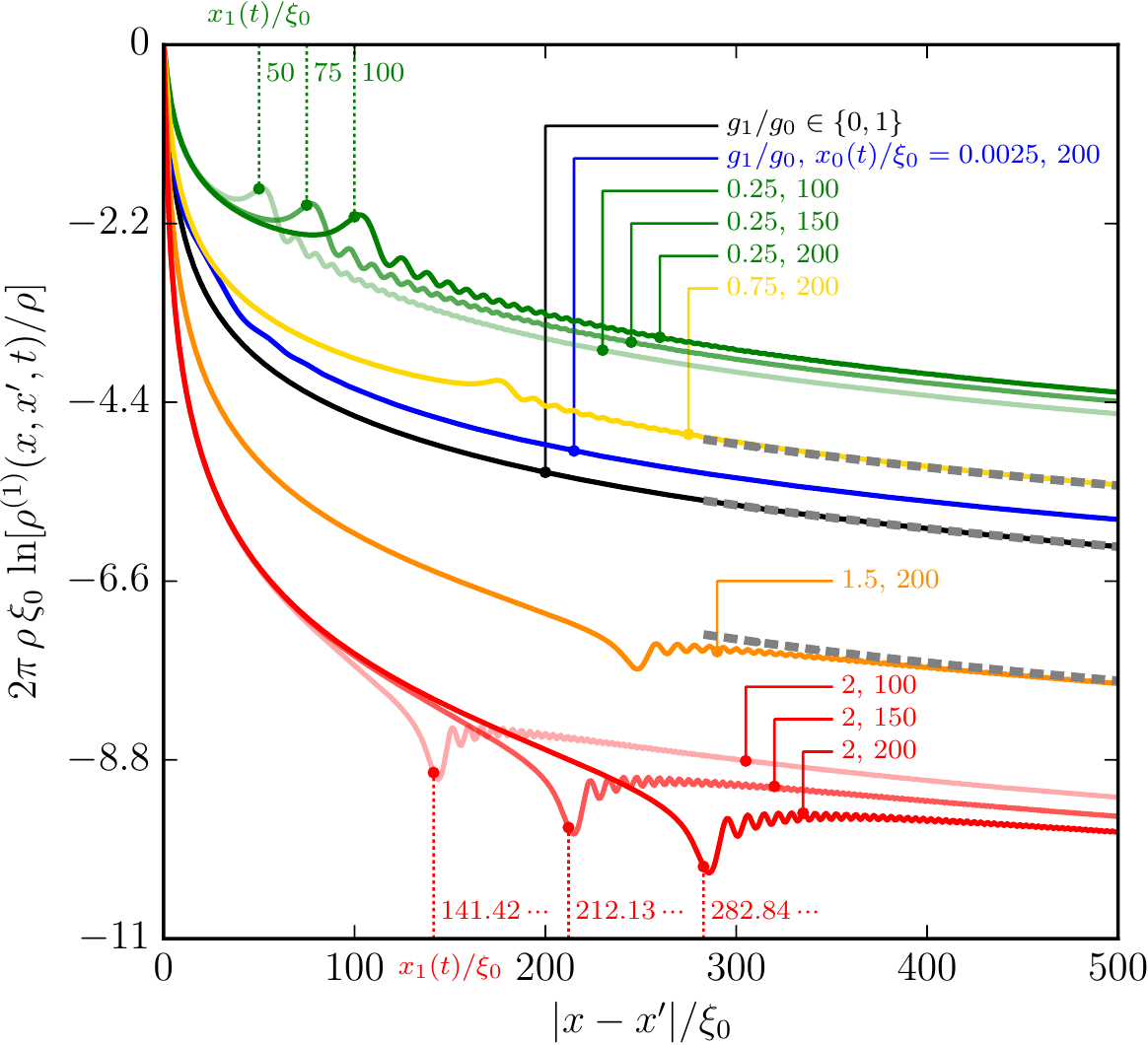}
\caption{Zero-temperature one-body density matrix $\rho^{(1)}(x,x',t)$
for a steplike coupling of the form~\eqref{eq:sl_gt} with $g_0 \neq 0$,
as given in Eq.~\eqref{eq:OBDMQuenchStep}. Different values of the
coupling ratio $g_1/g_0$ and of the dimensionless time $x_0(t)/\xi_0$
elapsed since $t=0$ are considered. The black curve, obtained for
$g_1 = 0$ or $g_0$, corresponds to the prequench or the no-quench
equilibrium result [first contribution on the right-hand side of
Eq.~\eqref{eq:OBDMQuenchStep}]. The gray dashed curves indicate the long-range
behavior~\eqref{eq:OBDMQuenchStepLongRange}--\eqref{eq:OBDMQuenchStepLongRangeParameter}
for $g_1/g_0 = 0$, 0.75, 1, and 1.5 at $x_0(t)/\xi_0 = 200$. The vertical
dotted lines indicate the dimensionless ``Lieb-Robinson bounds''
$x_1(t)/\xi_0$ [see Eq.~\eqref{eq:LRBound}] for $g_1/g_0 = 0.25$ (green)
and $g_1/g_0 = 2$ (red) at $x_0(t)/\xi_0 = 100$ (light color), $150$ (normal color),
and $200$ (dark color).}
\label{fig:OBDMQuenchStep}
\end{figure}

Let us first consider the system to be initially in its ground state
at $T = 0$. In this case, its Bogoliubov momentum distribution
$n_k(t) = n_k(t)|_0$ after the quench is explicitly given in
Eq.~\eqref{eq:sl_depl_T0_t0mInf_tpInf}, from which we infer the
following dimensionless expression for the corresponding one-body
density matrix:
\begin{align}
\notag
&\left.2\pi\,\rho\,\xi_0\,
\ln\!\bigg[\frac{\rho^{(1)}(x,x',t)}{\rho}\bigg]\right. \\
\notag
&\left.=-\int_0^{+\infty}dq\,\bigg(\frac{q^{2}+2}{q\,\sqrt{q^{2}+4}}-1\bigg)
\,\bigg[1-\cos\!\bigg(q\,\frac{|x-x'|}{\xi_0}\bigg)\bigg]\right. \\
\notag
&\left.\hphantom{=}+8\,\frac{g_1}{g_0}\,\bigg(1-\frac{g_1}{g_0}\bigg)
\int_0^{+\infty}dq\,\frac{\displaystyle{\sin^{2}\!
\bigg[\frac{q}{4}\,\sqrt{q^{2}+4\,\frac{g_1}{g_0}}\,
\frac{x_0(t)}{\xi_0}\bigg]}}
{\displaystyle{q\,\bigg(q^{2}+4\,\frac{g_1}{g_0}\bigg)\,
\sqrt{q^{2}+4}}}\right. \\
\label{eq:OBDMQuenchStep}
&\left.\hphantom{=}\times
\bigg[1-\cos\!\bigg(q\,\frac{|x-x'|}{\xi_0}\bigg)\bigg] \, . \right.
\end{align}
In this equation,
\begin{equation}
\frac{x_0(t)}{\xi_0} =
\frac{2 c_0 t}{\xi_0}=\frac{2 \mu_0 t}{\hbar}
\end{equation}
is the time elapsed after the quench in units of
$\hbar / (2 \mu_0)$, where $\mu_0 = g_0 \rho$ denotes
the chemical potential of the system before the quench. The
corresponding density matrix is represented in
Fig.~\ref{fig:OBDMQuenchStep} as a function of $|x-x'|/\xi_0$
for different values of $g_1/g_0$ and
$x_0(t)/\xi_0$.

\begin{figure*}
\includegraphics[width=0.7\linewidth]{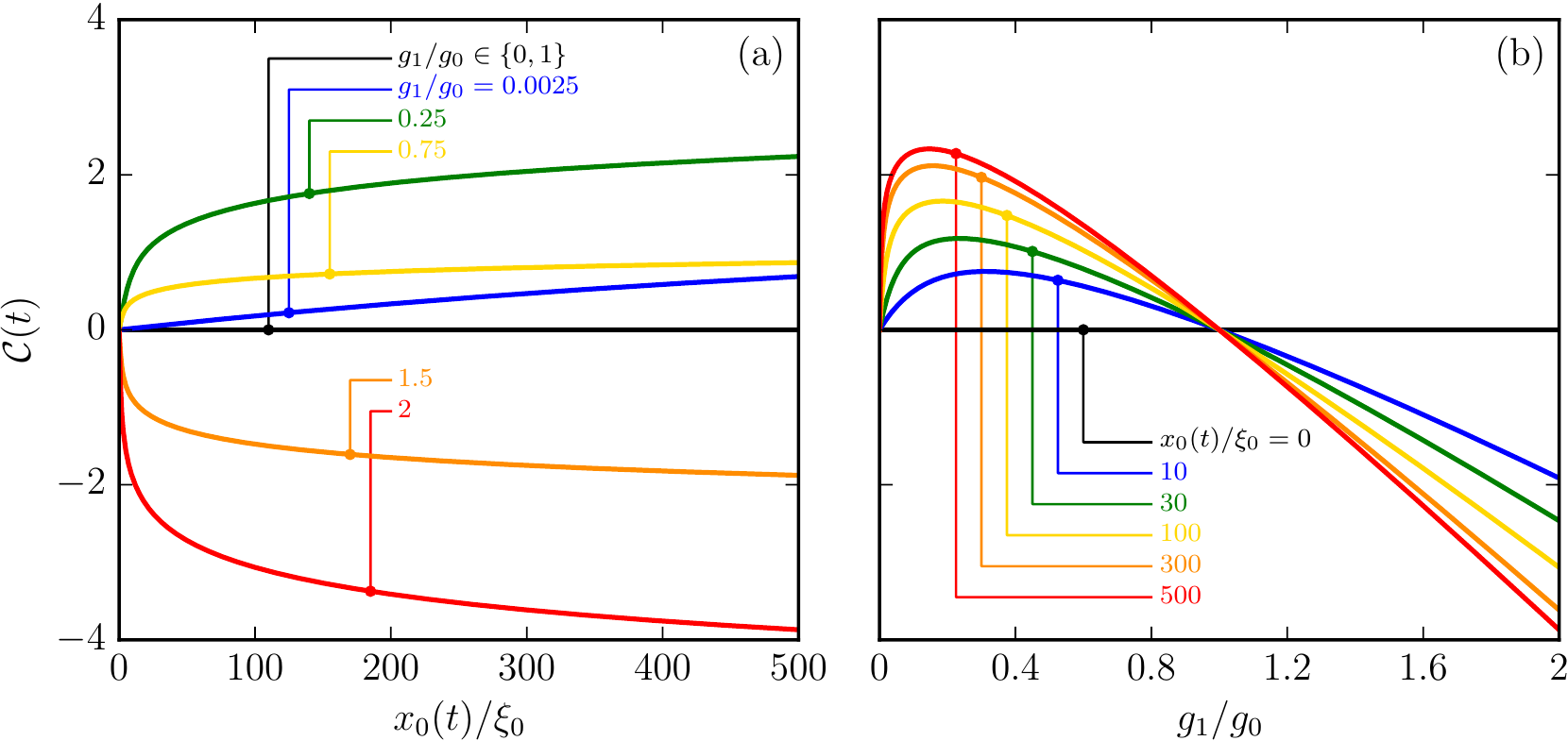}
\caption{Parameter $\mathcal{C}(t)$, as given by
Eq.~\eqref{eq:OBDMQuenchStepLongRangeParameter}, as a function of
the dimensionless time $x_0(t)/\xi_0$ and for different coupling ratios
$g_1/g_0$ [panel (a)], and as a function of $g_1/g_0$ for different
$x_0(t)/\xi_0$ [panel (b)]. For $g_1 = 0$ or $g_0$ (freezing
of the momentum distribution or no quench), one has $\mathcal{C}(t) = 0$
[black solid line on (a) and intersection points of the curves on (b)].}
\label{fig:Ct}
\end{figure*}

The first integral on the right-hand side of Eq.~\eqref{eq:OBDMQuenchStep}
gives the one-body density matrix of the not-yet-quenched system,
with constant coupling constant $g_0$~\cite{Mor03,Lar13}. The
quench-induced time dependence is embodied in the second
contribution. The latter is zero when $g_1 = 0$ or $g_0$
which is easily understandable. When $g_1 = 0$, the postquench
bosons no longer interact and their $\rho^{(1)}$ function
remains frozen to its initial shape. When $g_1 = g_0$, there
is no quench at all and the $\rho^{(1)}$ function is the
equilibrium one with coupling constant $g_0$ at all times. We
also understand the sign of this second contribution: When
$g_1 < g_0$ ($g_1 > g_0$), the quantum fluctuations
are globally reduced (increased) and coherence is accordingly
increased (reduced).

At short distances, when $|x-x'|$ is much smaller than
$\min\{\xi_0,\xi_1\}$, the $\rho^{(1)}$ function is
Gaussian, which is typical of noninteracting systems (see, e.g.,
Ref.~\cite{Nar99}). At large distances instead, when
$|x-x'|$ is much larger than the Lieb-Robinson--type bound
\begin{equation}
x_1(t) = 2 c_1 t \, ,
\label{eq:LRBound}
\end{equation}
$\rho^{(1)}$ decays as a power law:
\begin{equation}
\rho^{(1)}(x,x',t) \simeq \rho\,
\bigg(\frac{\exp[2-\gamma+\mathcal{C}(t)]}{4}\,
\frac{\xi_0}{|x-x'|}\bigg)^{\alpha} \, .
\label{eq:OBDMQuenchStepLongRange}
\end{equation}
In this expression, $\gamma=0.57(7)$ is the Euler-Mascheroni
constant and the exponent
\begin{equation}
\alpha = \frac{1}{2\pi\rho\xi_0}
\label{eq:alpha}
\end{equation}
is exactly the same as the one governing the long-range decay of the
prequench, equilibrium one-body density matrix (see, e.g.,
Refs.~\cite{Mor03,Lar13}). The time- and coupling-dependent
parameter
\begin{align}
\notag
\mathcal{C}(t)&\left.=8\,\frac{g_1}{g_0}\,
\bigg(1-\frac{g_1}{g_0}\bigg)\right. \\
\label{eq:OBDMQuenchStepLongRangeParameter}
&\left.\hphantom{=}\times\int_0^{+\infty}dq\,
\frac{\displaystyle{\sin^{2}\!\bigg[\frac{q}{4}\,
\sqrt{q^{2}+4\,\frac{g_1}{g_0}}\,\frac{x_0(t)}{\xi_0}\bigg]}}
{\displaystyle{q\,\bigg(q^{2}+4\,\frac{g_1}{g_0}\bigg)\,
\sqrt{q^{2}+4}}}\right.
\end{align}
embodies the long-distance effect of the quench. $\mathcal{C}(t)$ is
plotted as a function of $x_0(t)/\xi_0$ for different values of
$g_1/g_0$ in Fig.~\ref{fig:Ct}(a), and as a function of $g_1/g_0$ for
different values of $x_0(t)/\xi_0$ in Fig.~\ref{fig:Ct}(b). Note that
$\mathcal{C}(t=0) = 0$: In this case, the asymptotic
behavior~\eqref{eq:OBDMQuenchStepLongRange} of the one-body density
matrix is just the prequench one. As expected, $\mathcal{C}(t)$
cancels when $g_1 = 0$ or $g_0$ (freezing of the momentum
distribution or no quench, respectively); it is positive for
$g_1 < g_0$ (since in this case the quench leads to an increased
coherence) and negative for $g_1 > g_0$ (decreased coherence). One may
also evaluate its large-time [that is, $x_0(t)/\xi_0 \gg 1$] behavior
in both limits where $g_1$ is very small or very large compared to
$g_0$. When $g_1/g_0 \ll 1$, we replace $q^2 + 4 g_1/g_0$ with $q^2$
in the integrand of Eq.~\eqref{eq:OBDMQuenchStepLongRangeParameter}
and obtain
\begin{equation}
\mathcal{C}(t)\underset{t \to + \infty}{=}
\frac{\pi}{4}\,\frac{g_1}{g_0}\,\frac{x_0(t)}{\xi_0} \, .
\end{equation}
When $g_1/g_0 \gg 1$ instead, we replace $q^2 + 4 g_1/g_0$ with
$4 g_1/g_0$ in the integrand of
Eq.~\eqref{eq:OBDMQuenchStepLongRangeParameter} and get
\begin{equation}
\mathcal{C}(t)\underset{t \to + \infty}{=}
-\frac{1}{2}\,\frac{g_1}{g_0}\,
\bigg\{\ln\!\bigg[\sqrt{\frac{g_1}{g_0}}\,\frac{x_0(t)}{\xi_0}\bigg]
+2\ln2+\gamma\bigg\} \, .
\label{eq:LargeTCt}
\end{equation}
In the latter expression we kept subdominant contributions in order to
have a better approximation of the leading logarithmic term.

The quantity $x_1(t)$ introduced in Eq.~\eqref{eq:LRBound}
defines the boundary between the large-distance regime and what is
sometimes called the ``interior of the light cone.'' Its value is
indicated in Fig.~\ref{fig:OBDMQuenchStep} in units of $\xi_0$
for $g_1/g_0 = 0.25$ and $g_1/g_0 = 2$ at the dimensionless times
$x_0(t)/\xi_0 = 100$, $150$, and $200$. Dispersive effects are
important in our system and the speed of sound $c_1$ is not an exact
equivalent of the speed of transport of information. As can be seen
in Fig.~\ref{fig:OBDMQuenchStep}, this results in the fact that the
separation between the ``interior'' and the ``exterior of the light
cone'' is not sharp. A hint of the leaking of information ``outside of
the light cone'' is the fact that the parameter $\mathcal{C}(t)$
involved in the long-range behavior of $\rho^{(1)}$ depends on
$t$: the correlation between particles separated by a distance
exceeding the ``Lieb-Robinson bound'' $x_1(t)$ is affected by
the quench.

In the particular case where $g_0 = 0$ and $g_1 > 0$ (not shown in
Fig.~\ref{fig:OBDMQuenchStep}), one has a thermal-like exponential decrease
of $\rho^{(1)}$ ``within the light cone,'' irrespective of whether the
system is clean~\cite{Lar16} or (weakly) disordered~\cite{Lar18}.
No such prethermalization effect is observed in the generic situation
where $g_0 \neq 0$. Note also that, just beyond $x_1(t)$, the one-body
density matrix displays small-amplitude oscillations. The latter originate
from the high-momentum Bogoliubov excitations generated in response to the
sudden quench~\cite{Lar16}.

In Ref.~\cite{Sch18}, Schemmer \textit{et al.} consider a thermally
occupied initial state. This drastically modifies the long-range
behavior of $\rho^{(1)}$, which no longer decays as a power law but
exponentially, both before and after the quench. In the large-$|x-x'|$
limit, the integral in Eq.~\eqref{eq:OBDMQuench} is naturally
dominated by the infrared contribution. In this phonon limit, the
zero-temperature and thermal contributions to the Bogoliubov momentum
distribution~\eqref{eq:GenericMomentumDistribution} reduce to
\begin{align}
n_k(t)|_0&\simeq\frac{1}{2}\,
\frac{1+(g_1/g_0-1)\sin^{2}[k\,x_1(t)/2]}{|k|\,\xi_0}, \\
\label{eq:MomDistribT}
n_k(t)|_{T}&\simeq\frac{k_{\mathrm{B}}\,T}{\mu_0}\,
\frac{1+(g_1/g_0-1)\sin^{2}[k\,x_1(t)/2]}{(k\,\xi_0)^{2}} \, .
\end{align}
One thus sees that $n_k(t)|_{T}$ dominates over $n_k(t)|_0$ in the
$|k|\xi_0 \ll 1$ regime, which indicates that the long-range
$\rho^{(1)}$ function at finite temperature behaves differently from
its zero-temperature counterpart. The corresponding Bogoliubov momentum
distribution, which is approximately equal to~\eqref{eq:MomDistribT},
then may be cast in the form
\begin{equation}
\label{eq:MomDistrib}
n_k(t)\simeq\frac{k_{\mathrm{B}}\,T}{\mu_0}\,
\frac{1}{(k\,\xi_0)^{2}}+2\,\frac{k_{\mathrm{B}}\,T_{\ast}}
{\mu_0}\,\frac{\sin^{2}[k\,x_1(t)/2]}{(k\,\xi_0)^{2}} \, ,
\end{equation}
where we introduced the effective temperature
\begin{equation}
T_{\ast}=\frac{1}{2}\,\bigg(\frac{g_1}{g_0}-1\bigg)\,T \, .
\end{equation}
Inserting Eq.~\eqref{eq:MomDistrib} into Eq.~\eqref{eq:OBDMQuench} we
eventually obtain the following expression for the long-range
$\rho^{(1)}$ function:
\begin{equation}
\label{eq:Th}
\frac{\rho^{(1)}(x,x',t)}{\rho}\simeq
\exp\!\left[{-}\pi\,\frac{|x-x'|}{\rho\,\Lambda^{2}(T)}-
\pi\,\frac{|x-x'|}{\rho\,\Lambda^{2}(T_*)}\right]
\end{equation}
when $|x-x'|\leqslant x_1(t)$, and
\begin{equation}
\frac{\rho^{(1)}(x,x',t)}{\rho}\simeq
\exp\!\left[{-}\pi\,\frac{|x-x'|}{\rho\,\Lambda^{2}(T)}
-\pi\,\frac{x_1(t)}{\rho\,\Lambda^{2}(T_*)}\right]
\end{equation}
when $|x-x'|>x_1(t)$. In the two above equations
$\Lambda(T) = h(2\pi\,m\,k_{\mathrm{B}}\,T)^{-1/2}$
is the thermal de Broglie wavelength.

As a result, in the very-long-time limit $x_1(t)/\xi_0 \to +\infty$,
the long-range $\rho^{(1)}$ function essentially reaches the form
expected for a weakly interacting 1D thermal state,
\begin{equation}
\label{eq:OBDMSch}
\rho^{(1)}(x,x',t)\simeq
\rho\,\exp\!\bigg[{-}\pi\,
\frac{|x-x'|}{\rho\,\Lambda(T_{\mathrm{fin}})^{2}}\bigg] \, ,
\end{equation}
with a final temperature~\cite{Sch18}
\begin{equation}
T_{\mathrm{fin}} =
T+T_{\ast}=\frac{1}{2}\,\bigg(1+\frac{g_1}{g_0}\bigg)\,T \, .
\end{equation}
We note here that similar results have also been obtained in the
theoretical study of a quenched pair of one-dimensional Bose gases
within the Luttinger liquid approach~\cite{Lan18b}.

Note that with a $-\pi$ instead of a $-2\pi$ in the
argument of the exponential, the long-range, long-time thermal
one-body density matrix~\eqref{eq:OBDMSch} is very close to that of an
ideal gas. The difference is due to the fact that the phase fluctuations
are dominant over the density fluctuations in the 1D quasi-ideal regime
whereas they equally contribute in the ideal case
(see, e.g., Ref.~\cite{Bou11}).

\subsection{Woods-Saxon-type coupling in 1D}
\label{subsec:1DWS}
In this section, we consider a situation where the nonlinear coupling
constant $g(t)$ obeys a smooth temporal transition from 
$g_0 \neq 0$ to $g_1 \geq 0$ according to the
law~\eqref{eq:ws_gt}. We assume here that the system is initially
in its ground state at $T=0$.

\begin{figure}[t!]
\includegraphics[width=\linewidth]{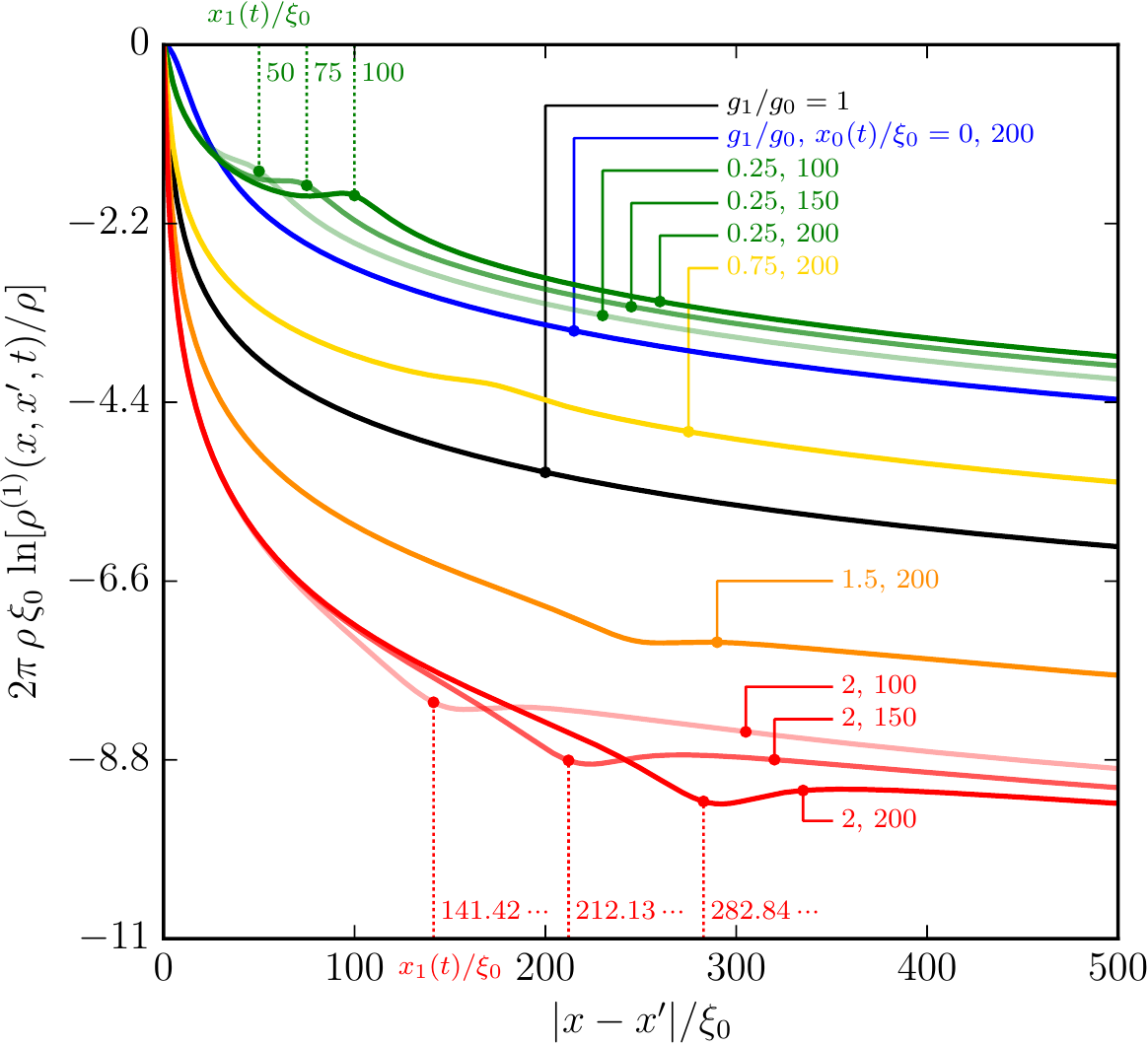}
\caption{Same as Fig.~\ref{fig:OBDMQuenchStep} for a Woods-Saxon-type
coupling of the form~\eqref{eq:ws_gt} with $x_0(\tau)/\xi_0 = 0$, as given in
Eqs.~\eqref{eq:WSRho1}. The black curve, obtained for $g_1 = g_0$,
corresponds to the no-quench equilibrium result.}
\label{fig:OBDMQuenchWoodsSaxon}
\end{figure}

From Eq.~\eqref{eq:OBDMQuench} and the zero-temperature Bogoliubov
momentum distribution~\eqref{eq:GenericMomentumDistribution-b} computed
using Eqs.~\eqref{eq:quad_uv_time} and~\eqref{eq:ws_gamma}, we obtain the
following expression for 
the one-body density matrix at some time $t \geq 0$:
\begin{widetext}
\begin{equation}
\label{eq:WSRho1}
\begin{split}
\notag
2\pi\,\rho\,\xi_0\,\ln\!\bigg[\frac{\rho^{(1)}(x,x',t)}{\rho}\bigg]&
\left.=-\frac{1}{2}\int_0^{+\infty}dq\;\bigg|
\frac{q-\sqrt{q^{2}+4}}{(q\,\sqrt{q^{2}+4})^{1/2}}\,
\frac{{}_{2}F_{1}[\alpha_q,
\gamma_q-\beta_q,\gamma_q,(1+e^{-t/\tau})^{-1}]}
{(1+e^{t/\tau})^{\alpha_q}}\right. \\
&\left.\hphantom{=}+2\,\bigg(\frac{\sqrt{q^{2}+4}}{q}\bigg)^{1/2}\,
\frac{\alpha_q\,\beta_q}{(\alpha_q+\beta_q)\,\gamma_q}\,
\frac{{}_{2}F_{1}[\alpha_q+1,\gamma_q-\beta_q,
\gamma_q+1,(1+\mathrm{e}^{-t/\tau})^{-1}]\,
e^{t/\tau}}{(1+e^{t/\tau})^{\alpha_q+1}}\bigg|^{2}\right. \\
&\left.\hphantom{=}\times\bigg[1-\cos\!\bigg(q\,\frac{|x-x'|}
{\xi_0}\bigg)\bigg] \, , \right.
\end{split}
\end{equation}
\end{widetext}
where, from Eqs.~\eqref{eq:ws_abc},
\begin{subequations}
\label{eq:OBDMQuenchWoodsSaxon}
\begin{align}
\label{eq:OBDMQuenchWoodsSaxon-c}
\alpha_q & \left.=-i\,
\frac{q}{4}\,\bigg(\sqrt{q^{2}+4}\,-\sqrt{q^{2}+4\,
\frac{g_1}{g_0}}\bigg)\,\frac{x_0(\tau)}{\xi_0} \, , \right. \\
\label{eq:OBDMQuenchWoodsSaxon-d}
\beta_q & \left.=-i\,\frac{q}{4}\,
\bigg(\sqrt{q^{2}+4}\,+\sqrt{q^{2}+4\,\frac{g_1}{g_0}}\bigg)\,
\frac{x_0(\tau)}{\xi_0} \, , \right. \\
\label{eq:OBDMQuenchWoodsSaxon-e}
\gamma_q & \left.=1-i\,\frac{q}{2}\,\sqrt{q^{2}+4}\;
\frac{x_0(\tau)}{\xi_0} \, . \right.
\end{align}
\end{subequations} 
The corresponding $\rho^{(1)}$ is plotted in
Fig.~\ref{fig:OBDMQuenchWoodsSaxon} as a function of
$|x-x'|/\xi_0$ for $x_0(\tau) / \xi_0 = 10$ and for several
values of $g_1 / g_0$ and $x_0(t) / \xi_0$.

As in the case of a steplike quench in $g(t)$, the long-range
$\rho^{(1)}$ function presents a power-law decay roughly of the
form~\eqref{eq:OBDMQuenchStepLongRange}. In the present case,
contrarily to what has been done in
Eq.~\eqref{eq:OBDMQuenchStepLongRangeParameter}, an analytic
expression for $\mathcal{C}(t)$ is not easily obtained, but one may
show by detailed numerical inspection that the exponent $\alpha$ which
governs the large-distance power-law decrease of the zero-temperature
$\rho^{(1)}$ is the same as the one given in Eq.~\eqref{eq:alpha},
which governs the long-range behavior of the prequench one-body
density matrix. Also in the present case, the regular and continuous
time dependence of the coupling parameter smooths out the oscillations
observed around the Lieb-Robinson bound $x_1(t)$ for the
steplike-quench one-body density matrix (compare
Figs.~\ref{fig:OBDMQuenchStep} and~\ref{fig:OBDMQuenchWoodsSaxon}).

Similar results have been obtained in Ref.~\cite{Ber14} in the
framework of Luttinger liquid description for a piecewise linear
function $g(t)$. In this reference, as in the present work, the
light-cone effect is less marked than for an abrupt quench and
accompanied by no oscillations in $\rho^{(1)}$. Here, however, at
variance with Ref.~\cite{Ber14}, we do not need to introduce an
effective Lieb-Robinson bound for a correct description of the
transition between short- and long-distance behavior.

\section{Conclusion}
\label{sec:conclusion}
We have analyzed some of the most relevant properties of a weakly
interacting uniform Bose gas having a time-varying coupling strength. 
In three dimensions the system can be considered as a condensate
(which can be described within a mean-field approach) with small
additional quantum fluctuations. These can be treated within a
time-dependent Bogoliubov framework, that enables one to determine the
time evolution of any observable. One gets a useful physical insight
into the problem by viewing each excited mode as a time-dependent
harmonic oscillator, whose frequency coincides with the instantaneous
Bogoliubov one. Using this correspondence, we prove some general
properties, such as the freezing of the low-momentum modes, the
adiabatic behavior of the high-momentum ones, and the possible
occurrence of Sakharov oscillations at large evolution times.  It
would be interesting to also use the gravitational analogy to study
temporal evolutions relevant in cosmology; work in this direction is
in progress.

Additionally, by mapping the problem onto a scattering one, we
identified a few families of time-dependent couplings whose evolution
equations can be solved analytically. For these models, we calculate
the quantum depletion (in three dimensions) and the full one-body density
matrix (in one dimension), that characterize the degree of
coherence of the system. In the one-dimensional case our results point
to the absence of prethermalization for a typical quench protocol when
the Bose gas is initially interacting and at zero temperature. On the other
hand, at finite initial temperature, the density matrix evolves to a
configuration typical for a new thermal state~\cite{Sch18}.

\begin{acknowledgments}
We thank D. Cl\'{e}ment, J. P. Corson, V. Fleurov, H. Landa, M. Mancini,
E. Orignac, D. Papoular, G. Roux, R. Santachiara, G. V. Shlyapnikov,
S. Stringari, and P. Zi\'{n} for fruitful discussions. This work was
supported by the French ANR under Grant No. ANR-15-CE30-0017 (Haralab
project) and by the Spanish Ministerio de Econom\'{i}a, Industria y
Competitividad Grants No. FIS2014-57387-C3-1-P and No. FIS2017-84440-C2-1-P,
the Generalitat Valenciana Project No. SEJI/2017/042 and the Severo Ochoa
Excellence Center Project No. SEV-2014-0398. The research leading to
these results has received funding from the European Research
Council under European Community's Seventh Framework Programme
(FP7/2007-2013 Grant Agreement No. 341197).
\end{acknowledgments}

\appendix
\section{Change of initial time}
\label{sec:change_initial_time}
Let us assume that we know the solution $\gamma_k(t,t_0)$ of the TDHO
equation~\eqref{eq:quad_tdho_eq_gamma} for a given initial time $t_0$,
and that we want to calculate $\gamma_k(t,t_0')$ with $t_0' > t_0$.
Evidently, for $t \leq t_0'$ one has $\gamma_k(t,t_0') = \exp[-i \omega_k(t_0')
(t-t_0')]$. For $t > t_0'$ one can express $\gamma_k(t,t_0')$ as a linear
combination of $\gamma_k(t,t_0)$ and $\gamma_k^*(t,t_0)$. The coefficients
of the combination are set by the requirement that $\gamma_k(t,t_0')$ and its
first-order derivative be continuous at $t=t_0'$. By doing this, one ends up
with the expression (valid for $t>t_0'$)
\begin{equation}
\begin{split}
\gamma_k(t,t_0') = {}&{} \sqrt{\frac{\omega_k(t_0')}{\omega_k(t_0)}} \\
{}&{} \hspace{-1cm} \times \left[ \beta_{1,k}^*(t'_0,t_0)  \gamma_k(t,t_0) 
- \beta_{2,k}(t_0',t_0) \gamma_k^*(t,t_0) \right] \, .
\end{split}
\label{eq:scatt_gamma_t0new}
\end{equation}
Here, the two quantities $\beta_{1,k}(t_0',t_0)$ and $\beta_{2,k}(t_0',t_0)$
are given by
\begin{subequations}
\label{eq:quad_beta_time_t0new}
\begin{align}
\beta_{1,k}(t_0',t_0) + \beta_{2,k}(t_0',t_0)
&{} = \sqrt{\frac{\omega_k(t_0')}{\omega_k(t_0)}} \, \gamma_k(t_0',t_0) \, ,
\label{eq:quad_betap_time_t0new} \\
\beta_{1,k}(t_0',t_0) - \beta_{2,k}(t_0',t_0)
&{} = \sqrt{\frac{\omega_k(t_0)}{\omega_k(t_0')}} \, \tilde{\gamma}_k(t_0',t_0) \, .
\label{eq:quad_betam_time_t0new}
\end{align}
\end{subequations}
They coincide with the entries of the quasiparticle
propagator~\eqref{eq:quasipart_evol_op} calculated at $t=t_0'$. This can be
verified, for instance, starting from Eqs.~\eqref{eq:uv_time} and expressing
$\beta_{1,k}(t,t_0)$ and $\beta_{2,k}(t,t_0)$ in terms of $U_k(t,t_0)$ and
$V_k(t,t_0)$. Then, combining the result with Eqs.~\eqref{eq:quad_uv_time}
and setting $t=t_0'$, one gets the relations~\eqref{eq:quad_beta_time_t0new}.

The large-$t$ behavior of $\gamma_k(t,t_0')$ can be found from that
of $\gamma_k(t,t_0)$ given by Eq.~\eqref{eq:scatt_gamma_tpInf}. One
finds
\begin{equation}
\begin{split}
\gamma_k(t,t_0') \underset{t \to + \infty}{=} {}&{}
e^{i \omega_k(t_0') t_0'}
\sqrt{\frac{\omega_k(t_0')}{\omega_k(+\infty)}} \\
{}&{} \hspace{-1cm} \times
\left[ \tl_k(t_0') e^{-i \omega_k(+\infty) t}
+ \tl_k(t_0') e^{i \omega_k(+\infty) t} \right] \, ,
\end{split}
\label{eq:scatt_gamma_t0new_tpInf}
\end{equation}
where the transfer coefficients at the new initial time $t_0'$ are
\begin{subequations}
\label{eq:scatt_tc_t0new}
\begin{align}
\begin{split}
\tl_k(t_0')
= {}&{} e^{i [\omega_k(t_0) t_0 - \omega_k(t_0') t_0']} \\
&{} \times \left[\beta_{1,k}^*(t_0',t_0) \tl_k(t_0)
- \beta_{2,k}(t_0',t_0) \tr_k(t_0)\right] \, ,
\end{split}
\label{eq:scatt_tc_left_t0new} \\
\begin{split}
\tr_k(t_0')
= {}&{} e^{i [\omega_k(t_0) t_0 - \omega_k(t_0') t_0']} \\
&{} \times \left[\beta_{1,k}^*(t_0',t_0) \tr_k(t_0)
- \beta_{2,k}(t_0',t_0) \tl_k(t_0)\right] \, .
\end{split}
\label{eq:scatt_tc_right_t0new}
\end{align}
\end{subequations}

\section{Proof of the constancy of the asymptotic 
value of the condensate depletion}
\label{sec:asymp_depl}
The starting point is the asymptotic expression~\eqref{eq:scatt_depl_T0_tpInf}
of the momentum distribution, which we rewrite expanding the square
modulus:
\begin{equation}
\begin{split}
n_{\vec{k}}(t) \underset{t \to + \infty}{=} {}& 
|v_k(+\infty) \tl_k(t_0)|^2 + |u_k(+\infty) \tr_k(t_0)|^2 \\
&{} + 2 u_k(+\infty) v_k(+\infty) \\
& \phantom{{}+} \times
\real\left[\tl_k(t_0) \trs_k(t_0) e^{- 2 i \omega_k(+\infty) t}\right] \, .
\end{split}
\label{eq:scatt_depl_T0_tpInf_expand}
\end{equation}
Integrating both sides of this equality over the whole momentum space
one reproduces Eq.~\eqref{eq:scatt_tot_depl_T0_tpInf}, provided that the integral
of the third term on the right-hand side vanishes for $t \to + \infty$.
This statement is trivially true if $g(+\infty) = 0$ because $v_k(+\infty) = 0$.
Therefore, we need to prove that it also remains valid when $g(+\infty) \neq 0$.
Let us then try to evaluate
\begin{equation}
\begin{split}
& \int d^3 k \, u_k(+\infty) 
v_k(+\infty) \tl_k(t_0) \trs_k(t_0) e^{- 2 i \omega_k(+\infty) t} \\
&{}= - 2\pi \int_0^{+\infty} \!\!\!\!\!\! d k \, 
k^2 \frac{g(+\infty)\rho}{\hbar\omega_k(+\infty)}
\tl_k(t_0) \trs_k(t_0) e^{- 2 i \omega_k(+\infty) t} \, .
\end{split}
\label{eq:scatt_tot_depl_T0_tpInf_integral}
\end{equation}
Here, we have performed the trivial integration over the solid angle
and we have used the explicit formulas~\eqref{eq:Bogo_uv} for the
instantaneous Bogoliubov weights. It is convenient to change the
integration variable to $\eta = \omega_k(+\infty) t$. The Bogoliubov
dispersion~\eqref{eq:Bogo_freq} can be easily inverted to express
$k$ as a function of $\eta / t$. We get
\begin{equation}
k = \xi^{-1}(+\infty) \sqrt{2[ (1 + \eta^2 / \tilde{t}^2)^{1/2} - 1]} \, ,
\label{eq:scatt_k_eta_t}
\end{equation}
where $\tilde{t} = g(+\infty)\rho t / \hbar$ is a dimensionless time
variable. Then, Eq.~\eqref{eq:scatt_tot_depl_T0_tpInf_integral} can
be rewritten as $- 2 \sqrt{2} \pi \xi^{-3}(+\infty)
\int_0^{+\infty} d\eta \, F(\eta,\tilde{t}) \, e^{- 2 i \eta}$, where
\begin{equation}
\begin{split}
F(\eta,\tilde{t}) = {}&{}
\frac{1}{(1 + \eta^2 / \tilde{t}^2)^{1/2}} 
\left[ \frac{(1 + \eta^2 / \tilde{t}^2)^{1/2}-1}{\tilde{t}^2} \right]^{1/2} \\
&{} \times 
\tl_{\eta / \tilde{t}}(t_0) 
\trs_{\eta / \tilde{t}}(t_0) \, .
\end{split}
\label{eq:scatt_integrand_eta_t}
\end{equation}
We notice that the first factor in the right-hand size of
Eq.~\eqref{eq:scatt_integrand_eta_t} tends to $1$ as
$\tilde{t} \to + \infty$, whereas the second one behaves like
$\eta / \sqrt{2} \tilde{t}^2$. It remains to understand what
happens to the transfer coefficients $\tl_{\eta / \tilde{t}}(t_0)$
and $\trs_{\eta / \tilde{t}}(t_0)$ in this limit. To this purpose,
we make use of the relations
\begin{subequations}
\label{eq:scatt_tc}
\begin{align}
\begin{split}
\tl_k(t_0) = {}&{} \lim_{t \to + \infty} \Bigg\{
e^{- i \omega_k(t_0) t_0} e^{i \omega_k(+\infty) t} \\
&{} \hspace*{-1.5cm} \times \frac{1}{2} \left[
\sqrt{\frac{\omega_k(+\infty)}{\omega_k(t_0)}} \,
\gamma_k(t,t_0)
+ \sqrt{\frac{\omega_k(t_0)}{\omega_k(+\infty)}} \,
\tilde{\gamma}_k(t,t_0) \right] \Bigg\} \, ,
\end{split}
\label{eq:scatt_tc_left} \\
\begin{split}
\tr_k(t_0)  = {}&{} \lim_{t \to + \infty} \Bigg\{
e^{- i \omega_k(t_0) t_0} e^{-i \omega_k(+\infty) t} \\
&{} \hspace*{-1.5cm} \times \frac{1}{2} \left[ 
\sqrt{\frac{\omega_k(+\infty)}{\omega_k(t_0)}} \,
\gamma_k(t,t_0)
- \sqrt{\frac{\omega_k(t_0)}{\omega_k(+\infty)}} \,
\tilde{\gamma}_k(t,t_0) \right] \Bigg\} \, .
\end{split}
\label{eq:scatt_tc_right} 
\end{align}
\end{subequations}
Equations~\eqref{eq:scatt_tc} follow from the asymptotic
expressions of $\gamma_k$ [see Eq.~\eqref{eq:scatt_gamma_tpInf}]
and $\tilde{\gamma}_k$ (easily deduced from that of $\gamma_k$).
In Sec.~\ref{subsec:freezing_low_mom} we have seen that,
if $g(t_0) \neq 0$, then $\gamma_k(t,t_0) \to 1$ and
$\tilde{\gamma}_k(t,t_0) \to 1$ for $k \to 0$. This yields
\begin{equation}
\tl_k(t_0) \trs_k(t_0) \underset{k \to 0}{=}
\frac{1}{4} \left[ \sqrt{\frac{g(+\infty)}{g(t_0)}}
- \sqrt{\frac{g(t_0)}{g(+\infty)}} \right] \, .
\label{eq:scatt_tc_lowk_g0f}
\end{equation}
Instead, if $g(t_0) = 0$, at low $k$ one has $\gamma_k(t,t_0) \to 1$,
$\tilde{\gamma}_k(t,t_0) \to Z(t,t_0)$ [with $Z$ given by
Eq.~\eqref{eq:pert_z}], and
\begin{equation}
\tl_k(t_0) \trs_k(t_0) \underset{k \to 0}{=}
\frac{m c(+\infty)}{2 \hbar k} \, .
\label{eq:scatt_tc_lowk_g00}
\end{equation}
Thus, we find that in the $\tilde{t} \to + \infty$ limit
the product $\tau^\leftarrow_{\eta / \tilde{t}}(t_0)
\tau^{\rightarrow *}_{\eta / \tilde{t}}(t_0)$ approaches a finite
value if $g(t_0) \neq 0$, while it behaves like $\tilde{t} / 2 \eta$
if $g(t_0) = 0$. In both cases $F(\eta,\tilde{t})$ vanishes
at large times, and so does its integral over $\eta$. Hence,
we conclude that the time-dependent terms in
Eq.~\eqref{eq:scatt_depl_T0_tpInf_expand} do not contribute
to the total depletion at $t \to + \infty$.

\section{Analytic continuation of hypergeometric functions}
\label{sec:hyperg_func}
The hypergeometric function is usually defined as the sum of the power
series~\cite{Abr65}
\begin{equation}
{}_{2}F_1(a,b,c;z) = \sum_{n=0}^{+\infty} \frac{(a)_n (b)_n}{(c)_n} 
\frac{z^n}{n!} \, ,
\label{eq:hyperg_def}
\end{equation}
where $(a)_n = \Gamma(a+n) / \Gamma(a)$ is the Pochhammer symbol.
This power series converges only if $|z| < 1$. However, the
hypergeometric function can be analytically continued along any path
in the complex plane that avoids the branch points $z=1$ and
$\infty$. For instance, in Sec.~\ref{subsec:ws_coupling}
we had to consider hypergeometric functions with real negative argument.
One can deal with such a situation by resorting to the Pfaff transformations
\begin{subequations}
\label{eq:hyperg_pfaff_trans}
\begin{align}
{}_{2}F_1(a,b,c;z) &{} = (1-z)^{-a} 
{}_{2}F_1(a,c-b,c;\textstyle\frac{z}{z-1}) \, ,
\label{eq:hyperg_pfaff_trans_1} \\
{}_{2}F_1(a,b,c;z) &{} = (1-z)^{-b} 
{}_{2}F_1(c-a,b,c;\textstyle\frac{z}{z-1}) \, .
\label{eq:hyperg_pfaff_trans_2}
\end{align}
\end{subequations}
These identities hold where the domains of the functions on both
sides overlap. If $|z| \geq 1$ and $\real z < 1/2$,
Eqs.~\eqref{eq:hyperg_pfaff_trans} define the analytic continuation
of the hypergeometric function to this latter domain, where the
right-hand side is well defined. An additional advantage of
Eqs.~\eqref{eq:hyperg_pfaff_trans} is that they provide a more efficient
way to compute numerically the hypergeometric function for negative argument.
Indeed, since $\left| \frac{z}{z-1} \right| < |z|$ when $z < 0$, the power
series expansions~\eqref{eq:hyperg_def}  of the functions on the right-hand
side converge faster than that of ${}_{2}F_1(a,b,c;z)$. The composition
of the two Pfaff transformations yields the Euler transformation
\begin{equation}
{}_{2}F_1(a,b,c;z) = (1-z)^{c-a-b} {}_{2}F_1(c-a,c-b,c;z) \, .
\label{eq:hyperg_euler_trans}
\end{equation}
There are many other identities in literature that allow one to
establish the analytic continuation of the hypergeometric
function~\cite{Abr65}. Among these we mention the following:
\begin{subequations}
\label{eq:hyperg_mInf}
\begin{align}
\begin{split}
{}_{2}F_1(a,b,c;z) = {}& 
\frac{\Gamma(c) \Gamma(b-a)}{\Gamma(b) \Gamma(c-a)}  (-z)^{-a} \\
& \times {}_{2}F_1(a,1-c+a,1-b+a;z^{-1}) \\
&{} + \frac{\Gamma(c) \Gamma(a-b)}{\Gamma(a) \Gamma(c-b)} (-z)^{-b} \\
& \times {}_{2}F_1(b,1-c+b,1-a+b;z^{-1}) \, ,
\end{split}
\label{eq:hyperg_mInf_1} \\
\begin{split}
{}_{2}F_1(a,b,c;z) = {}& 
\frac{\Gamma(c) \Gamma(c-a-b)}{\Gamma(c-a) \Gamma(c-b)} \\
& \times {}_{2}F_1(a,b,a+b-c+1;1-z) \\
&{} + \frac{\Gamma(c) \Gamma(a+b-c)}{\Gamma(a) \Gamma(b)} 
(1-z)^{c-a-b} z^{1-c} \\
& \times {}_{2}F_1(1-b,1-a,c-a-b+1;1-z) \, .
\end{split}
\label{eq:hyperg_mInf_2}
\end{align}
\end{subequations}
Equations~\eqref{eq:hyperg_mInf_1} and~\eqref{eq:hyperg_mInf_2}
are used in Secs.~\ref{subsec:ws_coupling} and~\ref{subsec:pt_coupling},
respectively, to analytically extend the solutions of the TDHO
equation for the Woods-Saxon and the modified P\"{o}schl-Teller
couplings. They are especially useful for extrapolating the large-$t$
behavior, yielding the transfer coefficients. Notice that, strictly speaking,
the domains of the hypergeometric functions on the two sides of the
identity~\eqref{eq:hyperg_mInf_1} do not overlap: the left-hand side is
defined for $|z|<1$, the right-hand side for $|z|>1$. In writing this formula
we implicitly assume that an intermediate analytic continuation around $z=-1$
is made, e.g., through the Pfaff transformations~\eqref{eq:hyperg_pfaff_trans}.
Concerning Eq.~\eqref{eq:hyperg_mInf_2}, both sides are simultaneously
defined for $0 < z < 1$, hence, there is no need for intermediate continuations.

\end{document}